\documentclass[A4 paper, 10pt]{elsarticle}

\usepackage{amsmath, amsfonts, amssymb}
\usepackage{graphicx}
\usepackage{epstopdf}
\usepackage{epsfig}
\usepackage[ruled, lined, linesnumbered]{algorithm2e}
\usepackage{setspace}
\usepackage{nomencl}
\usepackage{xcolor}

\usepackage{lipsum}
\makeatletter
\def\ps@pprintTitle{%
 \let\@oddhead\@empty
 \let\@evenhead\@empty
 \def\@oddfoot{}%
 \let\@evenfoot\@oddfoot}
\makeatother

\def\hybrid{\topmargin -20pt  \oddsidemargin 0pt
	\headheight 0pt \headsep 0pt
	\textwidth 6.25in	
	\textheight 9.5in	
	\marginparwidth .875in
	\parskip 5pt plus 1pt		\jot = 1.5ex}

\hybrid

\DeclareMathOperator{\expt}{\mathit{E}}

\DeclareMathOperator{\diag}{\mathit{diag}}

\DeclareMathOperator{\KL}{\mathit{KL}}
\DeclareMathOperator{\Ord}{\mathit{O}}
\DeclareMathOperator{\tr}{\mathit{tr}}
\DeclareMathOperator{\E}{\mathit{E}}
\DeclareMathOperator{\Var}{\mathit{Var}}
\newcommand{\dd}{\mathrm{d}}
\DeclareMathOperator{\func}{\mathit{f}}

\begin{document}
\begin{frontmatter}

\title{Information Geometry and Sequential Monte Carlo Samplers}
	\author[rvt]{Aaron Sim\corref{cor1}}
	\ead{aaron.sim11@imperial.ac.uk}
	\author[rvt]{Sarah Filippi}
	\ead{s.filippi@imperial.ac.uk}
	\author[rvt]{Michael P. H. Stumpf\corref{cor1}}
	\ead{m.stumpf@imperial.ac.uk}
	
	\cortext[cor1]{Corresponding authors}
	\address[rvt]{Centre for Integrative Systems Biology and Bioinformatics, Department of Life Sciences,\\ Imperial College London, UK, SW7 2AZ}

\begin{abstract}
This paper explores the application of methods from information geometry to the sequential Monte Carlo (SMC) sampler. In particular the Riemannian manifold Metropolis-adjusted Langevin algorithm (mMALA) is adapted for the transition kernels in SMC. Similar to its function in Markov chain Monte Carlo methods, the mMALA is a fully adaptable kernel which allows for efficient sampling of high-dimensional and highly correlated parameter spaces. We set up the theoretical framework for its use in SMC with a focus on the application to the problem of sequential Bayesian inference for dynamical systems as modelled by sets of ordinary differential equations. In addition, we argue that defining the sequence of distributions on geodesics optimises the effective sample sizes in the SMC run. We illustrate the application of the methodology by inferring the parameters of simulated Lotka-Volterra and Fitzhugh-Nagumo models. In particular we demonstrate that compared to employing a standard adaptive random walk kernel, the SMC sampler with an information geometric kernel design attains a higher level of statistical robustness in the inferred parameters of the dynamical systems.
\end{abstract}

\begin{keyword}
Information geometry \sep Sequential Monte Carlo \sep Bayesian inference \sep Dynamical Systems \sep mMALA
\end{keyword}

\end{frontmatter}

\section{Introduction}\label{introduction}
Sequential Monte Carlo (SMC) techniques are very much the sampler \textit{du jour} for a wide range of tasks traditionally tackled using Markov chain Monte Carlo (MCMC) methods  \cite{Doucet:2010ue,DelMoral:2006hc}. In the context of sequential Bayesian inference, for example, these include posterior sampling and model selection. The distinguishing characteristic of SMC sampling algorithms, as the name suggests, is a sequence of intermediate distributions $\{\pi_1(x_1), \pi_2(x_2), \dotsc \}$. In early applications of SMC methods (e.g. particle filters), such sequences would typically consist of non-homogeneous distributions defined on supports of different, often increasing, dimensions (i.e. $\dim(\mathcal{X}_{a}) > \dim(\mathcal{X}_{b})$ for $a> b$, where $x_a\in \mathcal{X}_a$). Also, the algorithm outputs are usually samples from every distribution in the sequence. In more recent incarnations of SMC algorithms, in particular those viewed as alternatives to MCMC methods, these intermediate distributions only play a collective supporting role of bridging the gap between a tractable initial distribution and a target distribution of interest; it is also often the case, for example in tempered sequences, that the entire sequence is defined on a single support.

It is the presence of a common support which, together with the freedom of constructing the set of intermediate distributions, throws up an important design consideration for optimising the efficiency of the algorithm. The entire sequence, book-ended by the initial and target distributions, now defines a path in the space of distributions. As a proxy to the task of minimising the cumulative distances between the distributions in the chain, one seeks the `shortest' continuous route between the initial and target distributions. This is not unlike the requirement in sequential importance sampling (SIS) for adjacent importance distributions to be `similar' to each other. The overall aim of this paper is to explore two aspects of this design guidance. The first aspect is the identity of the sequence itself, while the second is the perturbation methods used to move along this pre-specified sequence of distributions. Apart from several trivial examples where an analytical solution for the shortest path, or geodesic, exists (e.g. sampling from a multivariate normal distribution), consideration of this first aspect is usually restricted to the constrained problem of selecting the optimal spacing of distributions along a given, possibly arbitrarily chosen, path. For example, in a sequence of $p$ tempered distributions 
\begin{equation}
\bigl\{\pi_a(x) \sim \exp \bigl(\phi_a \func(x)\bigr)\bigr\}_{a\in\mathbb{T}},\label{tempereddef}
\end{equation}
where $\mathbb{T}=\{1,2,\dotsc, p\}$, and $\func$ some specified parametric function of $x$, selecting the optimal sequence simply involves tuning and selecting an optimal sequence of parameters $\{\phi_a\}_{a \in \mathbb{T}}$. 

The specification of geodesics, together with the concept of movement in distribution-space strongly suggests that adopting a geometrical angle may provide a unifying framework for the consideration of the above design issues. Indeed such a geometrical framework was first introduced to MCMC in \cite{Girolami:2011hw} where the authors employed ideas from information geometry to the design of efficient MCMC kernels and mooted its extension to population Monte Carlo (PMC) methods. Other recent examples of statistical applications of Riemannian geometry are its use in increasing the efficiency of Variational Bayesian methods \cite{honkela} and in sensitivity analysis in stochastic models \cite{Komorowski:2011cn}, amongst others.

Far from being an independent approach, SMC, together with other PMC methods, is often described as \textit{parallel} MCMC, in part because the sampled variables, or particles, are typically perturbed along the sequences of intermediate distributions using MCMC proposals and are accepted or rejected based on the standard evaluation of the Metropolis-Hastings (MH) ratio. Seen in this context it is not surprising for theoretical developments in MCMC algorithms, specifically in the design of efficient kernel proposals, to find their way into SMC methods. In fact, as it has been demonstrated in \cite{BenCalderhead:2011vx}, the well-documented advantage of PMC over MCMC samplers in addressing the issue of multiple distribution modes is preserved when adopting the differential geometric kernels. This paper follows this trend where we extend to the SMC sampler of Del Moral \textit{et al} \cite{DelMoral:2006hc} the work of Girolami and Calderhead \cite{Girolami:2011hw}.

The elegance of a geometrical framework is very often the sole justification for its construction. However, in the case of the SMC sampler, the benefits  extend beyond mere aesthetics; geometrically optimal placements of distributions and perturbation kernels will result in improved particle acceptance rates with greater effective sample sizes, which in turn lead to more robust and, hence, reliable sampling statistics over the course of the algorithm. 

The rest of this paper is organised as follows. In section \ref{preliminaries} we provide a brief review of the theoretical background to our work -- namely SMC samplers and the relevant aspects of information geometry. In section \ref{IGSMC} we consider geodesics on statistical manifolds and adapt the differential geometric MCMC kernels for use in SMC samplers. We illustrate the latter application to sequential Bayesian inference by way of three examples. We start with a trivial example involving the univariate Gaussian distribution and follow up with simulations involving the Fitzhugh-Nagumo and Lotka-Volterra ordinary differential equation (ODE) models. We conclude in chapter \ref{discussion} with a summary and discussion.

\paragraph{Note on notation} We have adopted the Einstein summation convention. Also, components of a metric $g$ and its inverse $g^{-1}$ are written with covariant ($g_{ij}$) and contravariant indices ($g^{ij}$) respectively, such that $g^{ik}g_{kj}=\delta^i_j$, with $\delta^i_j$ the Kronecker delta. Where there is no ambiguity, derivatives are abbreviated in the usual form as $\partial_i \equiv \frac{\partial}{\partial \xi^i}$ for a given variable $\xi^i$. Component-wise vector multiplication which maps $\mathbb{R}^n \times \mathbb{R}^n \rightarrow \mathbb{R}^n$ is specified by the $*$ symbol, where $(\mathbf{u}*\mathbf{v})_i \equiv u_iv_i$ (no sum), with $u_i$ and $v_i$ the $i$th components of vectors $\mathbf{u}$ and $\mathbf{v}$ respectively.

\section{Theoretical background}\label{preliminaries}

\subsection{Sequential Monte Carlo Samplers}\label{SMCsamp}
There are, at present, several different implementations of SMC algorithms (see, for instance, \cite{Doucet:2010ue, Liu:2008ww}). With regards to the ability to admit intermediate distributions defined on a common space -- a necessary requirement of our attempt to build a geometrical framework -- and the flexibility to define general moves between populations, it is the SMC sampler implementation by Del Moral \textit{et al} \cite{DelMoral:2006hc} which proves the most natural. We adopt this version in this paper and, in accordance with its authors, refer to it throughout simply as the \textit{SMC sampler}.

Similar to SIS, the SMC sampler is used to sample successively from a sequence of distributions. Unlike SIS, however, the SMC sampler does not require one to calculate the intermediate importance distributions. This is achieved through the introduction of artificial backward-in-time Markov kernels $L_a$ alongside the usual forward-in-time versions $K_a$,  the effect of which is a reduction of the complexity from $\Ord(N^2)$ to $\Ord(N)$, where $N$ is the number of sampled particles.

Let $\pi_p$, $p\in\mathbb{Z}$, be a target distribution for sampling. One specifies a preceding sequence of distributions $\{\pi_1, \pi_2, \dotsc, \pi_{p-1}\}$, with all $\{\pi_a\}_{a\in\mathbb{T}}$ defined on the support of $\pi_p$. Let $\{\eta_a\}_{a\in\mathbb{T}}$ represent the sequence of importance distributions where only the initial importance distribution $\eta_1$ is explicitly specified; the most natural choice is $\eta_1 = \pi_1$. In practice, the normalisation constants are unknown and one works instead with the unnormalised sequence $\{\gamma_a\}_{a\in \mathbb{T}}$, where 
\begin{equation}
\pi_a = \frac{\gamma_a}{Z_a}, \qquad\quad Z_a\in \mathbb{R}.
\end{equation}
One samples a population of $N$ particles $\{\xi_{a}^{(n)}\}_{n\in\{1,2,\dotsc, N\}}$ from the importance distribution $\eta_{a}$ by perturbing particles from the previous population via the transition kernels $K_a(\xi_{a}^{(n)}\, |\, \xi_{a-1}^{(n)})$ -- e.g. local random walks, Gibbs moves, etc. In this paper we use MCMC kernels for these local moves, the advantage being that the guaranteed convergence to each $\pi_a$ allows for several simplifying approximations to be made in the algorithm (see \cite{DelMoral:2006hc}). In particular, one has a simple approximation of a near-optimal expression of $L_a$ in terms of $K_a$ and $\pi_a$, which, in turn, simplifies the sequential updating of the particle weights -- i.e. the incremental factor to update the weights of a particle $\xi^{(n)}_{a-1}$ is
\begin{equation}
\tilde{w}_a(\xi_{a-1}^{(n)}, \xi_a^{(n)}) = \frac{\gamma_a(\xi_{a-1}^{(n)})}{\gamma_{a-1}(\xi_{a-1}^{(n)})}.
\end{equation}

The SMC sampler is summarised below (Algorithm \ref{SMCalgo}). For more details of the construction of the algorithm and proofs of consistency, we refer the reader to \cite{DelMoral:2006hc}.

\begin{algorithm}[H]
\KwIn{No. of particles per population $N$, Sequence of distributions $\pi_a$, $a=1,\dotsc, p$, Effective sample size (ESS) threshold $T$.}
\KwOut{Sampled particles $\xi^i_p$, Particle weights $W^i_p$,  $i=1,\dotsc, N$.}
Initialise $a=1$, ESS = $N$\;
Sample particles $\xi_1^{(n)}$ from prior  $\pi_1$\;
Set weights $W_1^{(n)} = \frac{1}{N}$\;
\For{$a\leq p$}{
  \If{$ESS < T$} {
    Resample particles $\{\xi_{a-1}^{(n)}\}$ from weighted multinomial distribution $\{(\xi_{a-1}^{(n)}, W_{a-1}^{(n)})\}$\;
    Reset weights $W_{a-1}^{(n)} = \frac{1}{N}$, $\forall i=1,\dotsc,N$\;  
  }
  \For{ $n = 1,2,\dotsc, N$}{
	Draw $\xi_a^{(n)} \sim K_a(\cdot\mid\xi^{(n)}_{a-1})$, where $K_a$ is a MCMC kernel\;
	Evaluate incremental weight $\tilde{w}_a(\xi_{a-1}^{(n)}, \xi_a^{(n)}) = \frac{\gamma_a(\xi_{a-1}^{(n)})}{\gamma_{a-1}(\xi_{a-1}^{(n)})}$\;
	Update particle weight $\widetilde{W}_a^{n} = W_{a-1}^{n}\cdot\tilde{w}_a(\xi_{a-1}^{(n)}, \xi_{a}^{(n)})$\;
  }
  Normalise particle weights $W_a^{n} = \widetilde{W}_a^{n}/\sum_{m=1}^{N} \widetilde{W}_a^{m}$\;
  Calculate $\text{ESS} = 1/\sum_{n=1}^N |W_a^{n}|^2$\;
  Set a = a+1;
  }
Return $\{\xi_p^{(n)}$, $W_p^n\}_{i\in \{1,\dotsc, N\}}$.

\caption{The SMC sampler with MCMC kernels}
\label{SMCalgo}
\end{algorithm}

For the remainder of the paper, where there is no ambiguity, we drop the particle index and simply write $\xi^{(n)}_a \rightarrow \xi_a$.

\subsection{Information geometry}\label{informationgeometry}
The premise of all of information geometry is the observation that the space of probability distributions on a set $\mathcal{X}$ can be viewed as a Riemannian manifold with a unique metric $g$ \cite{Amari:2007wc}. Clearly, this is an infinite-dimensional manifold. In most applications, however, one considers the projection to a finite-dimensional submanifold $\mathcal{S}$ by restricting to a specific statistical model of choice as represented by a set $\Xi$ of parameters. By expressing the set of coordinate functions on the manifold in terms of model parameters $\xi \in \Xi$, it is easy to conceptualise the manifold structure of $\mathcal{S}$. For example, the space of two-dimensional Gaussian distributions $\mathcal{N}(\mathbf{\mu}, \Sigma)$ is a five-dimensional manifold with (non-canonical) coordinate functions $\{\mu_1, \mu_2, \Sigma_{11}, \Sigma_{12}, \Sigma_{22}\}$.

The coordinate-specific metric $g_{ij}(\xi)$ on $\mathcal{S}=\{p(x; \xi)\mid\xi\in\Xi\}$ is defined by the expected Fisher information matrix
\begin{equation}
g_{ij}(\xi) := \expt_\xi[\partial_i\ell_\xi\partial_j\ell_\xi] = \int \partial_i\ell(x; \xi) \partial_j\ell(x; \xi)p(x; \xi) \, \dd x,
\end{equation}
where $\ell(x;\xi) \equiv \log p(x; \xi)$, and $\expt_\xi$ is the expectation with respect to the probability distribution $p(x; \xi)$. Using the property $\int p(x; \xi) \, dx=1$, one derives the useful alternative form of the Fisher metric
\begin{equation}
g_{ij}(\xi) = -\expt_\xi[\partial_i\partial_j\ell_\xi].\label{fisheralt}
\end{equation}

Apart from the invariance of $g_{ij}$ under a sufficient statistic map, the foundations of information geometry contain few other constraints. Specifically there are no restrictions on the class of  statistical models or on the identity of the measure space $\mathcal{X}$, which can be discrete or continuous, thereby explaining its wide applicability across many different fields. For example, in systems biology contexts, we often have
\begin{equation}
\mathcal{X}={(\mathbb{R}^+)^{T_1}\times (\mathbb{R}^+)^{T_2}\times \dotsb \times (\mathbb{R}^+)^{T_S}},
\end{equation}
where each sample $x\in \mathcal{X}$ represents a set of non-negative measurements (species abundance, chemical concentration, etc) of $S$ separate biological entities, each measured at $T_s$ separate time-points $(s=1,\dotsc,S)$.

With the metric $g_{ij}$, one proceeds to import the concepts and structures of Riemannian geometry like distances, geodesics, curvature, etc to a wide range of statistical analyses (see \cite{Amari:2007wc} for a concise introduction to the relevant aspects of Riemannian geometry). For example, we can define parallel transport on the space of distributions using the Levi-Civita connection $\nabla_{\partial_i}\partial_j = \Gamma^{k}_{\hphantom{k}ij} \partial_k$, where the Christoffel symbols $\Gamma$ are expressed, in the usual way, in terms of the Fisher metric and its derivatives as
\begin{equation}
\Gamma^{k}_{\hphantom{k}ij} = -\frac{1}{2}g^{kl}\bigl(\partial_i g_{jl} + \partial_j g_{il} - \partial_l g_{ij}\bigr).
\end{equation}
Using this expression we have the geodesics $\gamma: \mathbb{R} \rightarrow \mathcal{S}$ on $\mathcal{S}$ as solutions to the geodesic equation
\begin{equation}
\ddot{\gamma}^k(t) + \dot{\gamma}^i(t)\dot{\gamma}^j(t)( \Gamma^{k}_{\hphantom{k}ij})_{\gamma(t)}=0, \label{geodesiceom}
\end{equation}
where the dot refers to differentiation w.r.t. $t\in\mathbb{R}$, and the indices indicate the $i$th component in the given coordinate parameterisation, i.e. $\gamma^i(t) := \xi^i(\gamma(t))$, etc. 

In information geometry, the infinitesimal distance between two distributions is directly related to the Kullback-Leibler (KL) divergence $KL(\cdot\, |\, \cdot)$. With $\dd s^2 = g_{ij} \dd \xi^i\dd\xi^j$, we have \cite{Kullback:1968tm}
\begin{equation}
\KL\bigl(p(x; \xi + \dd \xi)\mid p(x; \xi) \bigr)=\frac{1}{2}\dd s^2.\label{KLrelation}
\end{equation}

\section{Information geometric design of SMC samplers}\label{IGSMC}

\subsection{Geodesics on statistical manifolds}\label{geodesicsonsm}

In many implementations of the SMC sampler, the sampling of each intermediate distribution is optimised by minimising the KL-divergence from the previous distribution in the sequence. This is often represented symbolically as $\pi_a\approx \pi_{a+1}$, where $a=1,2,\dotsc, p$.  Together with the relation in \eqref{KLrelation}, this suggests that, in the asymptotic limit $p \rightarrow \infty$, the full SMC sampler is optimised by selecting the intermediate distributions to lie on the geodesic connecting the fixed initial and final target distribution boundary points. We illustrate this by way of the following trivial example.

\paragraph{Sampling from a univariate Gaussian distribution} 
Assuming that one has a process of sampling from the standard normal $\mathcal{N}(0,1)$, one can use the SMC sampler to sample from an arbitrary distribution $\mathcal{N}(\mu', \sigma'^2)$. We consider sequences of 25 distributions on each of three separate paths on the univariate Gaussian manifold\footnote{Note that no attempt was made, on either geodesic or non-geodesic paths, to ensure equal Fisher distance between all pairs $\int_a^{a+1}ds$, as was done in \cite{Costa:2012wv}. Should we choose to space the distributions evenly, we fully expect the relative performance of the three paths, and our conclusions, to remain unchanged.}, with one of the paths being the geodesic. The analytic solution to the geodesic equation \eqref{geodesiceom} with fixed boundary points is provided in \ref{geoderive}. We compare the optimality of the SMC sampler by tracking the effective sample size as the SMC sampler progresses through the intermediate distributions. 

In the coordinate representation $(\mu, \sigma^2)$, we select arbitrary initial and target distributions $p_1 = (0,1)$ and $p_{25} = (5, 3)$ respectively. We employ a local random walk kernel with a fixed uniform proposal. Now in the SMC sampler algorithm (Algorithm \ref{SMCalgo}), the expression for the incremental weights $\tilde{w}_a$ is a good approximation for MCMC kernels only if the successive populations are close, i.e. $\pi_{a-1} \approx \pi_{a}$ (see \cite{DelMoral:2006hc} eq. 31). However, given that the effect of altering the distances between distributions is what we are aiming to test, there may be instances where this assumption is no longer valid. In its place, therefore, we use the full expression (\cite{DelMoral:2006hc} eq. 26) and particle approximation
\begin{equation}
\tilde{w}_a(\xi_{a-1}, \xi_a) =\frac{\gamma_a(\xi_a)}{\int_{\text{all } \zeta}\gamma_{a-1}(\zeta)K_a(\xi_{a}\mid \zeta)} \approx \frac{\gamma_a(\xi_a)}{\sum_{n=1}^{N}W^n_{a-1}K_a(\xi^{(n)}_{a}\mid \xi^{(n)}_{a-1})},\label{speker}
\end{equation}
where the full analytic expression for the kernel density $K_a(\xi^{(n)}_{a}\mid \xi^{(n)}_{a-1})$ is given in \ref{kernelderive}.

\begin{figure}[h] 
\centering
{\includegraphics{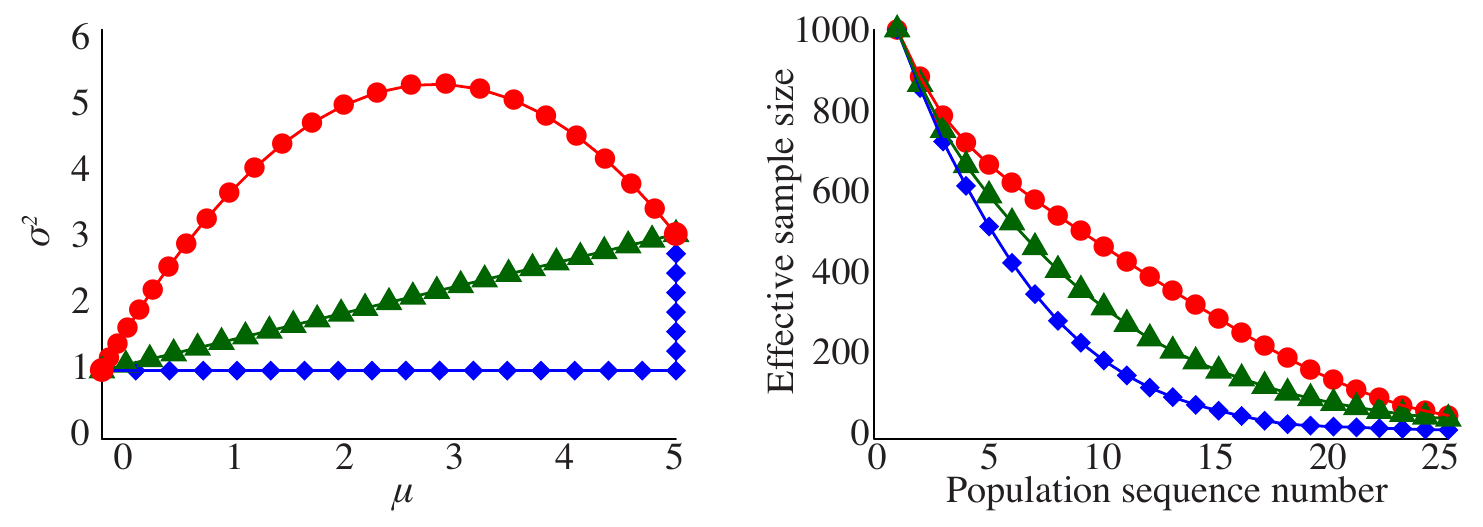}}
\caption[Comparison of geodesic paths]{\textbf{Left:} Sample paths between initial distribution $\mathcal{N}(0,1)$ and the target distribution $\mathcal{N}(5,3)$, each with 25 intermediate distributions (including end-distributions). \textbf{Right:} Evolution of ESS over 25 populations. The circular points (red) represent the distributions on the geodesic between the fixed boundaries; the triangles (green) indicate the naive straight-line path which implicitly (and erroneously) assumes a Euclidean metric; the diamonds (blue) indicate a two-staged path with the $\mu$ coordinate the first to be fully adjusted.}
\label{geodesicpath}
\end{figure}

Without resampling, we track the average progression of the mean effective sample sizes (ESS) across ten independent iterations of the SMC sampler. The results, together with the paths on the manifold, are shown in Figure \ref{geodesicpath}. It is clear that selecting the distributions to lie on the geodesics leads to a slower decline of the ESS over the course of the SMC sampler.

In practice, one would simply transform the samples from the standard normal $x \rightarrow x\sigma' + \mu'$. However, the availability of analytical solutions to the geodesic equation for Gaussian distributions presents an opportunity to employ the SMC sampler and to observe the effect of locating distributions on geodesics. Unfortunately, such analytical solutions are almost always unavailable. Indeed, as mentioned in the introduction, the procedure of writing down a tempered sequence \eqref{tempereddef} simply side-steps the issue by adopting the possibly non-optimal solution of fixing all but one of the parameters to the target distribution values. Effectively, this is equivalent to taking a one-dimensional submanifold, with parametrising coordinate $\phi$, in place of the true geodesic, and then attempting to atone for our sins by selecting a sequence of $\{\phi_a\}_{a\in\mathbb{T}}$ which minimises the cumulative distances for that particular path. One can also adopt a brute force approach to reducing the distance between each intermediate distribution by simply selecting a large number of intermediate distributions.

\subsection{Differential geometric MCMC kernels for SMC transitions}\label{mmalasec}
A non-optimal placement of intermediate distributions in SMC can be mitigated somewhat by employing an efficient kernel for transitions between distributions. In this section we describe one such method -- the manifold Metropolis-adjusted Langevin algorithm (mMALA) \cite{Girolami:2011hw}. mMALA is an algorithm which prescribes, using the local geometry of the statistical manifold, the discretised flow of a particle on a given space, whereby the sampled positions of the particle over time follow a specified distribution.

The mMALA MCMC kernel combines a discretised Brownian motion with drift proposal with the standard Metropolis-Hasting acceptance step. Let $\mathbf{\xi}\in\mathbb{R}^D$ be a random vector parametrising distributions $p(x; \xi)$, $x\in\mathcal{X}$. As described above, we treat $\xi$ as the coordinate functions of a $D$-dimensional manifold $\mathcal{S}$. Define $\tilde{p}(\xi)$ to be a density\footnote{Note that this density is an additional structure on the manifold; $\tilde{p}(\xi)$ is defined on $\mathcal{S}$, whereas for $s \in \mathcal{S}$, $p(x; \xi(s))$ is defined on $\mathcal{X}$.} on $\mathcal{S}$. In the special case where the target distribution is proportional to the likelihood of the set of $S$ observed samples $\{x_s\}$ from $\mathcal{X}$, i.e. $\tilde{p}(\xi) \propto \prod_{s=1}^{S}p(x_s;\xi)$, one can calculate the Fisher information matrix $g$ and track the discretised Langevin diffusion on the Riemannian manifold $(\mathcal{S}, g)$. As shown in \cite{Girolami:2011hw} The proposal density is the Gaussian
\begin{equation}\label{proposal}
q(\xi_{\tau+1}\mid\xi_\tau)\sim\mathcal{N}\bigl(\mu(\xi_\tau, \epsilon), \epsilon^2g^{-1}\bigr),
\end{equation}  
with $\epsilon$ the Langevin diffusion discretisation step size, $\tau$ the discrete time index, and where the components of the deterministic drift are given by
\begin{equation}\label{mu}
\mu^i(\xi_\tau, \epsilon) = (\xi_{\tau})^i +\frac{\epsilon^2}{2}\bigl(g^{ij}(\partial_j\ell(\xi_\tau))\bigr) - \epsilon^2\bigl(g^{ik}(\partial_jg_{kl})g^{lj}\bigr)+ \frac{\epsilon^2}{2}\bigl(g^{ij}g^{kl}(\partial_jg_{kl})\bigr),
\end{equation}
where $\ell(\xi) \equiv \log \tilde{p}(\xi)$.

The geometric design of efficient transition kernels for SMC is easily adapted from the MCMC context. Consider a sequence $\{\tilde{p}_a(\xi)\}_{a\in\mathbb{T}}$, defined on a common space $\mathcal{S}$. Where in MCMC a single metric $g_{ij}(\xi)$ is defined from the MCMC invariant density, here we have a sequence of metrics $\{(g_a)_{ij}\}_{a\in\mathbb{T}}$. Concomitantly, the set of proposal densities \eqref{proposal} is replaced by a sequence of proposal densities
\begin{equation}
\Bigl\{q_a(\xi_{a+1}\mid\xi_a) = \mathcal{N}\bigr(\mu_a(\xi_a, \epsilon), \epsilon^2g_a^{-1}\bigr)\Bigr\}_{a\in\mathbb{T}},\label{eulerprop}
\end{equation}
where $\mu_a(\xi_a, \epsilon)$ is given by \eqref{mu} with $g$ replaced, in sequence, by the metrics in $\{g_a\}_{a\in\mathbb{T}}$. 

Assuming that $\tilde{p}_a \approx \tilde{p}_{a+1}$ we expect a relatively high acceptance rate in the MH step. We can, therefore, get away with performing relatively fewer MCMC iterations (perhaps even just one per distribution in the sequence). Although this reduces the overall computational cost of the algorithm, it is dependent on the quality of the mixing property of the kernel, the latter of which can be improved, albeit at the expense of a decrease in acceptance rate, by choosing a larger Langevin diffusion step size $\epsilon$.

Apart from the choice of the initial distribution, there are two key algorithmic parameters to set: the discretisation step size, $\epsilon$, and the number of populations, $p$. We demonstrate the application of the mMALA---SMC combination and observe the implications of various parameter combination choices by way of the following simple example.

\subsection{Example: Univariate Gaussian parameter inference}\label{uniparam}
Consider $S$ observations, $x_i$, drawn from a univariate normal distribution $\mathcal{N}(\mu, \sigma^2)$ with parameters $\mu$ and $\sigma$ to be inferred using the SMC sampler. Adopting $\xi = (\xi_1, \xi_2) = (\mu, \sigma)$ as the coordinate functions of a manifold $\mathcal{S}$, we assume the normal priors $\pi_1(\xi_1) = \mathcal{N}(u_1, v_1^2)$ and $\pi_2(\xi_2) = \mathcal{N}(u_2, v_2^2)$ for some $u_1, v_1, u_2, v_2 \in \mathbb{R}$. We define a sequence of $p$ tempered distributions \cite{Neal:1996gr} $\{\tilde{p}_a(\xi)\}_{a\in \mathbb{T}}$, $\mathbb{T} = \{1,\dotsc, p\}$, where
\begin{equation}
\tilde{p}_a(\xi) = \pi(\xi)\biggl(\prod_{s=1}^{S}\frac{1}{\sqrt{2\pi \xi_2^2}}\exp -\frac{(x_s-\xi_1)^2}{2\xi_2^2}\biggr)^{\phi_a},\label{uninorm}
\end{equation}
with $0=\phi_1 < \phi_2 < \dotsm < \phi_{p-1} < \phi_{p} = 1$, and $\pi(\xi) = \diag(\pi_1(\xi_1), \pi_2(\xi_2))$. It can be shown that for tempered distributions of the exponential form (e.g. \eqref{uninorm}), the sum of symmetrised Kullback-Leibler divergences between the distributions is minimised by adopting a geometric tempering sequence where $\phi_{a+1}/\phi_{a} = \text{const}$ \cite{tempered}. Therefore, in the examples that follow, we adopt the geometric sequence with
\begin{equation}
\phi_a = \phi_2^{-\frac{a - 2}{p-2} +1},\quad \text{and} \quad\phi_1 = 0.\label{geometric}
\end{equation}
In addition, in this example, we fix $\phi_2 = 5\times 10^{-4}$.

Using \eqref{fisheralt} the metric for each distribution in the sequence is given by
\begin{equation}
g_a(\xi) = 
\begin{pmatrix} \frac{1}{v_1^2} + \frac{S\phi_a}{\xi_2^2} & 0 \\ 0 & \frac{1}{v_2^2} + \frac{2S\phi_a}{\xi_2^2}\\ \end{pmatrix}.\label{unimetric}
\end{equation}
For step-size $\epsilon$, the mMALA drift and the covariance matrix of the diffusion term on $\mathcal{S}$ are, respectively,
\begin{align}
\begin{pmatrix} \mu_\mu \\ \mu_\sigma \end{pmatrix} &= \frac{\epsilon^2}{2}
\begin{pmatrix}
\frac{v_1^2\xi_2^2}{C_1}\biggl[-\frac{\xi_1-u_1}{v_1^2} + \sum_{s=1}^{S}\frac{\phi_a(x_s-\xi_1)}{\xi_2^2}\biggr]\\
\frac{v_2^2\xi_2^2}{C_2}\biggl[-\frac{\xi_2-u_2}{v_2^2} - \frac{S\phi_1}{\xi_2} + \sum_{s=1}^S\frac{\phi_a (x_s - \xi_1)^2}{\xi_2^3} + \frac{2S\phi_a}{\xi_2}\Bigl(\frac{v_2^2}{C_2} - \frac{v_1^2}{2C_1}\Bigr)\biggr]
\end{pmatrix},\\
\Sigma_{ij}&\equiv \epsilon^2 g^{ij}= \epsilon^2
\begin{pmatrix} 
\frac{v_1^2\xi_2^2}{C_1} & 0 \\
0 & \frac{v_2^2\xi_2^2}{C_2}
\end{pmatrix},
\end{align}
with the $C_1 \equiv \xi_2^2 + v_1^2S\phi_a$ and $C_2 \equiv \xi_2^2 + 2v_2^2S\phi_a$. An important role played by the prior scale parameters $v_1$ and $v_2$ is that of setting a lower bound on the components of the metric $g_a$. The implications of this effect are examined in Section \ref{dysec}.

\begin{figure}[h] 
\centering
{\includegraphics{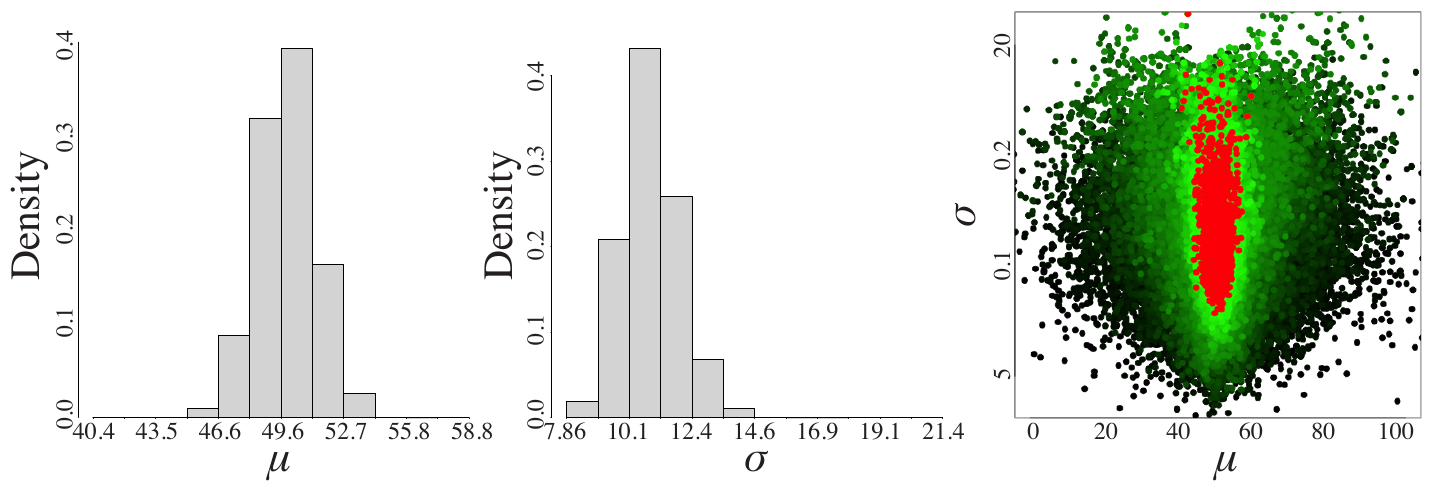}}
\caption[Marginal and joint distributions for univariate model parameter inference]{\textbf{Left/middle:} Weighted histogram of particles representing the estimated marginal distributions of the mean and standard deviation parameters.  \textbf{Right:} Scatterplot of particles over 45 populations for a simulated univariate model $\mathcal{N}(50, 10^2)$.  The initial particles are sampled from the prior $\pi(\xi) \equiv p_0(\xi) \sim \diag(\mathcal{N}(50, 20^2), \mathcal{N}(10, 2.5^2))$ and are coloured black in the scatter plot. The particles of the intermediate populations are represented in increasingly brighter shades of green with particles of the final population, representing the posterior, coloured red.}
\label{unijoint}
\end{figure}

\paragraph{Parameter inference}
We sample 60 points from the distribution $\mathcal{N}(\mu', \sigma'^2) = \mathcal{N}(50, 10^2)$ and let the prior distribution be centred on $(\mu', \sigma')$, i.e. $u_1 = 50$ and $u_2 = 10$, with $v_1 = 20, v_2 = 2.5$. Selecting a sequence of 45 populations and using 1500 particles, we run the SMC sampler with $\epsilon = 0.4$ and resampling fraction $T=0.3$. The estimates of the joint and marginal posteriors of $\mu, \sigma$ are presented in Figure \ref{unijoint}. As expected, the SMC sampler with mMALA transitions lead to a posterior which is centred around $(\mu, \sigma)=(50,10)$.

\begin{figure}[h] 
\centering
{\includegraphics{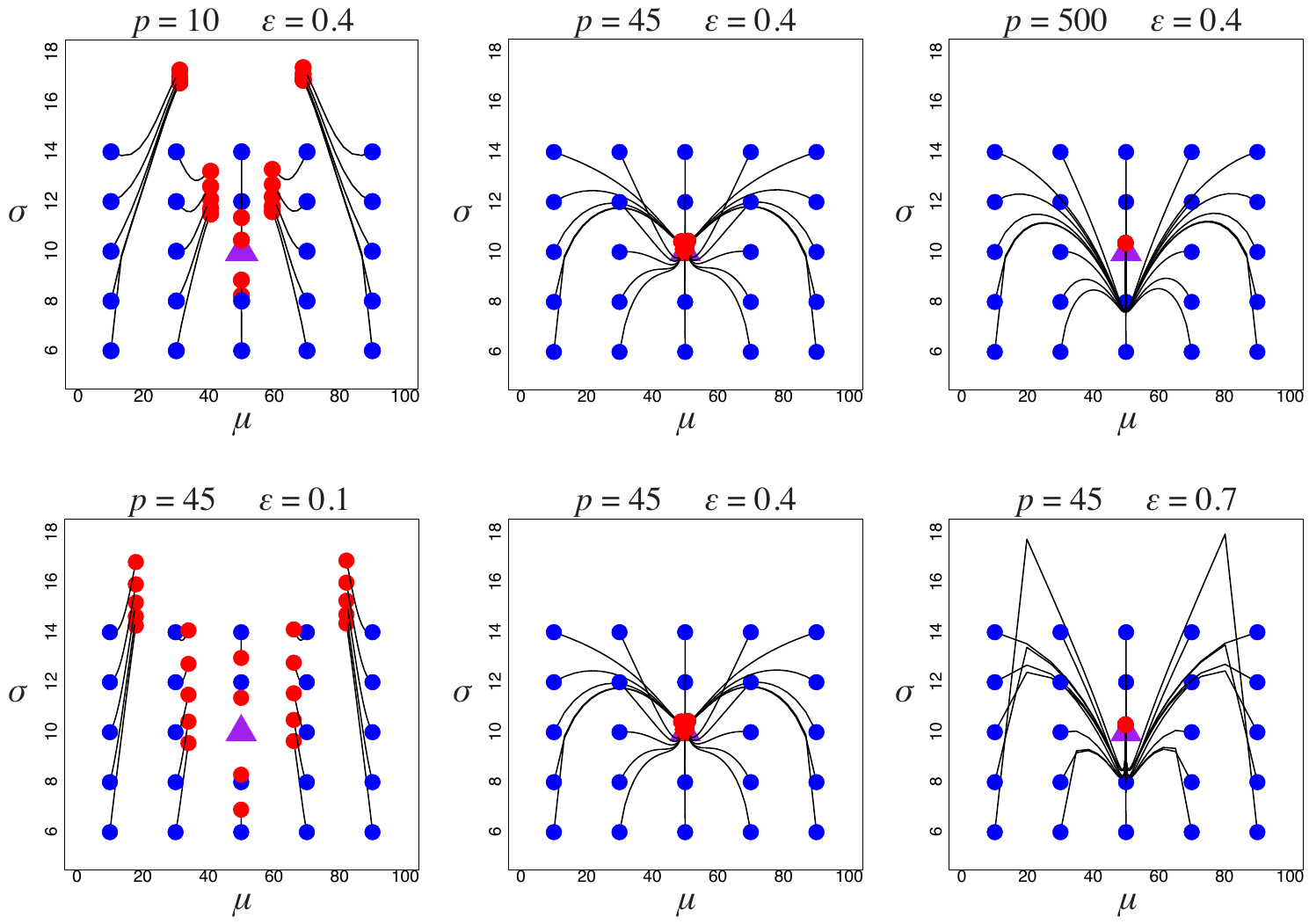}}
\caption[mMALA drift for univariate model -- effect of population number]{mMALA particle drift paths for the univariate normal model for different numbers of intermediate populations $p$ (top row), and step sizes $\epsilon$ (bottom row). The three population sizes and step sizes simulated are $p=10, 45, 500$, and $\epsilon=0.1, 0.4, 0.7$ respectively. The \textbf{purple} triangle marks the simulated mean and standard deviation $(\mu', \sigma') = (50, 10)$; the \textbf{blue} points represent the initial 24 particle samples and the \textbf{red} points the perturbed particles at the final population.}
\label{driftobs}
\end{figure}

\paragraph{mMALA drift}
In order to visualise the path of the particles and to understand the effects of varying the parameters $p$ and $\epsilon$ on the effectiveness of the SMC sampler, we focus on the deterministic perturbation at each intermediate stage by setting the diffusion term to zero and accepting all particles without resampling. We select 24 points on the manifold $\mathcal{S}$ from a grid surrounding the sample and observe the drift \eqref{mu} for varying numbers of intermediate populations $p$ and step-sizes $\epsilon$. The plots of the paths in $S$ are given in Figure \ref{driftobs}. We can already deduce several features of the mMALA drift. For a fixed step-size, $\epsilon$, a minimum acceptable number of populations for the SMC run with $p_{\text{min}}\sim 45$. If $p < 45$, the drift is too weak and the kernels do not mix well; if $p\geq45$, the particles do indeed drift toward the target coordinates with the variance in particle end-points decreasing as $p$ increases. If $\epsilon$ is too small, the drift is too weak, resulting in a poor performance of the SMC sampler; if $\epsilon$ is too large, the sharp changes in drift direction is an indication of the expected breakdown in the first-order Euler discretisation of the Langevin diffusion. The implication of this last point for SMC samplers is that a higher proportion of particles will be rejected in the MH-step leading to a lower number of effective particles used in the sampler. These results support the intuitive expectation that there is not just a minimum step size and total number of perturbation steps for effective sampling, but also a negative impact on the efficiency of the sampler when these tuning parameters are too large.

The results from this simple example hints at a strategy for optimising the differential geometric kernels of the SMC sampler: keeping constant the product of the square of the step-size and population number, i.e.
\begin{equation}
\epsilon^2p \sim \text{const},\label{const}
\end{equation}
we seek to minimise the number of populations whilst ensuring that the step-size is not so large that the precision of the algorithm is impacted by an increased rejection rate. For example, we found that the drift behaviour with $\epsilon = 0.4$, $p=45$ is essentially identical to the set-up with $\epsilon = 0.2$, and $p=180$ ($0.4^2\times 45 = 0.2^2\times 180$). Without factoring in the particle acceptance rate, the efficiency of the algorithm simply scales with the number of populations.

Other than the tuneable parameters, $p$ and $\epsilon$, it is the identity of the starting distribution, e.g. the prior in the context of Sequential Bayesian inference, which has a big impact on the viability of the mMALA transitions for SMC samplers. We defer this discussion to the following section where the implications are particularly significant.

\subsection{Application to parameter inference in dynamical systems} \label{dysec}
Following \cite{Girolami:2011hw}, we consider a system of differential equations
\begin{equation}
\dot{\mathbf{x}}(t) = f(\mathbf{x}, \xi, t),\label{de}
\end{equation}
where $\mathbf{x}$ is a $D$-dimensional vector, $\xi$ the model parameter vector, and $t$ the time variable. The observations $Y$ of $\mathbf{x}$ are made of the underlying state at $\tau$ time-points $\{t_1, t_2, \dotsc, t_\tau\}$ with a known observation noise model, i.e.
\begin{equation}
Y = X + E,
\end{equation}
where $X = X(\mathbf{x}_0, \xi) =  \bigl(\mathbf{x}(t_1)|\mathbf{x}(t_2)|\dotsm|\mathbf{x}(t_\tau)\bigr)^T$ is a solution to \eqref{de}, and $E$ the observation noise. $Y$, $X$, and $E$ are $(\tau\times D)$-dimensional matrices. For given initial conditions $\mathbf{x}_0$ and prior $\pi(\xi)$, the task is to estimate the posterior 
\begin{equation}
p(\xi\,|\,Y, \mathbf{x}_0, E)\propto \pi(\xi)\cdot L(Y\,|\,\xi, \mathbf{x}_0, E),\label{posterior}
\end{equation}
where $L(\cdot\,|\,\cdot)$ is the likelihood function. For example, as described in \cite{Girolami:2011hw}, for a noise model given by a time-independent normal distribution with variance $\sigma_d^2$, $d\in\{1,\dotsc, D\}$, we have
\begin{equation}
 L(Y\,|\,\xi, \mathbf{x}_0, E) = \prod_d^D\mathcal{N}\bigl(X(\xi, \mathbf{x}_0)_{\cdot, d}, \Sigma_d\bigr),\label{normL}
\end{equation}
where $\Sigma_d = \mathbb{I}_\tau\sigma_d^2$, and $X_{\cdot,d}$ denotes the time-series vector for species $d$. For a log-normal noise model with corresponding parameters, we have
\begin{equation}
 L(Y\,|\,\xi, \mathbf{x}_0, E) = \prod_d^D\log\mathcal{N}\bigl(X(\xi, \mathbf{x}_0)_{\cdot, d}, \Sigma_d\bigr).\label{logL}
\end{equation}

Observation noise in realistic dynamical systems are, however, likely to be more complicated. The only consequence of deviating from the above vanilla noise models is an added complexity in the Fisher metric calculations. As an example of the additional algebra involved, we have examined the implications of two such modifications -- heteroscedasticity and a truncation of the normal noise model. The details of these Fisher metric calculations are collected in \ref{mmaladev}. Nevertheless, since there are no conceptual novelties arising from cases with such adjustments, we have simply adopted the normal model of \eqref{normL} for the remainder of this section.

The problem can now be given a geometrical construction. We treat the model parameters $\xi$ as coordinate functions on a manifold $\mathcal{S}$. The combination of the differential equations \eqref{de} and an observation noise model allows us to associate to each point in $\mathcal{S}$ a probability distribution as follows. Define the measure space
\begin{equation}
\mathcal{X}\cong (\underset{D \text{ terms}}{\underbrace{\mathbb{R}^\tau\times \mathbb{R}^\tau\times \dotsb \times \mathbb{R}^\tau}}).\label{measuresp}
\end{equation}
The density $p(Y_{\cdot, n}; \xi)$ is given by the posterior in \eqref{posterior}, i.e.
\begin{equation}
p(Y_{\cdot, n}; \xi) =\kappa\pi(\xi)\cdot L(Y_{\cdot,n}\,|\,\xi, \mathbf{x}_0, E),\label{key}
\end{equation}
where $\kappa\in \mathbb{R}$ is a constant. Using this density we can then proceed to define the Fisher metric $g_{ij}(\xi)$ on $\mathcal{S}$ via \eqref{fisheralt}. For both the Gaussian and log-normal noise model, the likelihood has an exponential form. Hence, abbreviating $p(x; \xi) \equiv \kappa \pi(\xi)\exp\Phi(\xi)$, we have
\begin{equation}\label{integrand}
\begin{split}
-\partial_i\partial_j\bigl(\log p(Y_{\cdot,n}; \xi)\bigr) &= -\partial_i\partial_j\Phi - \frac{\bigl(\partial_i\partial_j\pi(\xi)\bigr)}{\pi(\xi)} + \frac{\bigl(\partial_i\pi(\xi)\bigr)\bigl(\partial_j\pi(\xi)\bigr)}{\pi^2(\xi)}\\
&\equiv -\partial_i\partial_j\Phi + h_{ij}(\xi),
\end{split}
\end{equation}
where $h_{ij}(\xi)$ depends only on $\xi$ via the prior $\pi(\xi)$ and not on $Y_{\cdot,n}$. We evaluate $h_{ij}(\xi)$ for several typical priors -- uniform, multivariate normal (MVN), and the component-wise (CW) log-normal -- as\footnote{Note: the indices are not summed over in the expression for the log-normal prior.}
\begin{align}
\text{Uniform:}&\qquad h_{ij}(\xi)= 0,\label{hij1}\\
\text{MVN:}&\qquad h_{ij}(\xi)= \bigl(\Sigma^{-1}\bigr)_{ij},\label{hij2}\\
\text{CW log-normal:}&\qquad h_{ij}(\xi)= \frac{\delta_{ij}}{(\xi^i\sigma^{i})^2}\bigl(1-\log \xi^i + \mu^i\bigr),\label{hij3}
\end{align}
where the $\mu_i$ and $\Sigma_{ij}$ are usual components of the mean vector and covariance matrix in $\dim(\mathcal{S})$-dimensions; we also assume that for the uniform prior, $h_{ij}(\xi)$ is evaluated away from the boundary of the non-zero support where $\partial_i\pi(\xi)$ is undefined.

In the framework of SMC we now define, on the shared measure space \eqref{measuresp}, the sequence of $p$ distributions
\begin{equation}
p_a(x; \xi) = \kappa\pi(\xi)\cdot \bigl[L(Y\,|\,\xi, \mathbf{x}_0, E)\bigr]^{\phi_a},\label{temperedds}
\end{equation}
with $a\in \mathbb{T}$ and $0=\phi_1 < \phi_2 < \dotsm < \phi_p =1$. From this sequence of distributions we have a matching sequence of Riemannian manifolds $\{(\mathcal{S}_a, g_a)\}_{a\in\mathbb{T}}$. Similar to the evaluation in \cite{Girolami:2011hw}, the Fisher metrics $(g_a)_{ij}$ and their first and second derivatives w.r.t. $\xi$, evaluated using \eqref{fisheralt}, can be written for the normal noise model \eqref{normL} as
\begin{align}
(g_a)_{ij} &= {\phi_a}\sum_{d=1}^{D}S_{i,d}^T\Sigma^{-1}_dS_{j,d} + \expt_{\xi}(h_{ij}) ,\label{gs}\\
\begin{split}
\partial_k(g_a)_{ij} &= {\phi_a} \sum_{d=1}^{D}\Bigl[(\partial_k S_{i,d})^T\Sigma^{-1}_dS_{j,d} + S_{i,d}^T\Sigma^{-1}_d(\partial_kS_{j,d})\Bigr] + \partial_k\bigl(\expt_{\xi}(h_{ij})\bigr),\label{dgs}
\end{split}
\end{align}
where the sensitivity $S$ is defined as
\begin{equation}
S_{i,d} := \frac{\dd X_d}{\dd\xi^i}.
\end{equation}
Details of the derivation of \eqref{gs} are given in \ref{mmaladev}. Following \cite{Girolami:2011hw}, we evaluate $S$ and its partial derivatives for all sampled time points via the numerical solutions to a set of auxillary differential equations obtained by repeatedly differentiating \eqref{de} w.r.t. $\xi$. These auxiliary equations are collected in \ref{sensitivityDE}.

We now explore the application of this methodology with two examples: the Fitzhugh-Nagumo and the Lotka-Volterra model.

\paragraph{Example: Fitzhugh-Nagumo model}
The Fitzhugh-Nagumo model was developed as a simplification of the Hodgkin-Huxley model, the latter describing the dynamics of the potentials of a spiking neuron. The dynamics of the voltage $V$ and response $R$ variables is modelled by the following set of ODEs
\begin{align}
\frac{\dd V}{\dd t} &= c\Bigl(V- \frac{V^3}{3} + R\Bigr),\\
\frac{\dd R}{\dd t} & = \frac{a-V-bR}{c},
\end{align}
where $a, b, c$ are the parameters of the system. Following the treatment in \cite{Girolami:2011hw} and \cite{Ramsay:2007gj}, we simulate the system with starting values $(V_0, R_0)=(-1,1)$ and parameters $(a', b', c')=(0.2, 0.2, 3)$. Assuming a fixed, identical (over all time-points), normal observation noise model for both potentials with $\sigma^2=0.05$, we make 25 observations and attempt to infer the parameter values using an SMC sampler with mMALA transitions. 

\begin{figure}[h] 
\centering
{\includegraphics{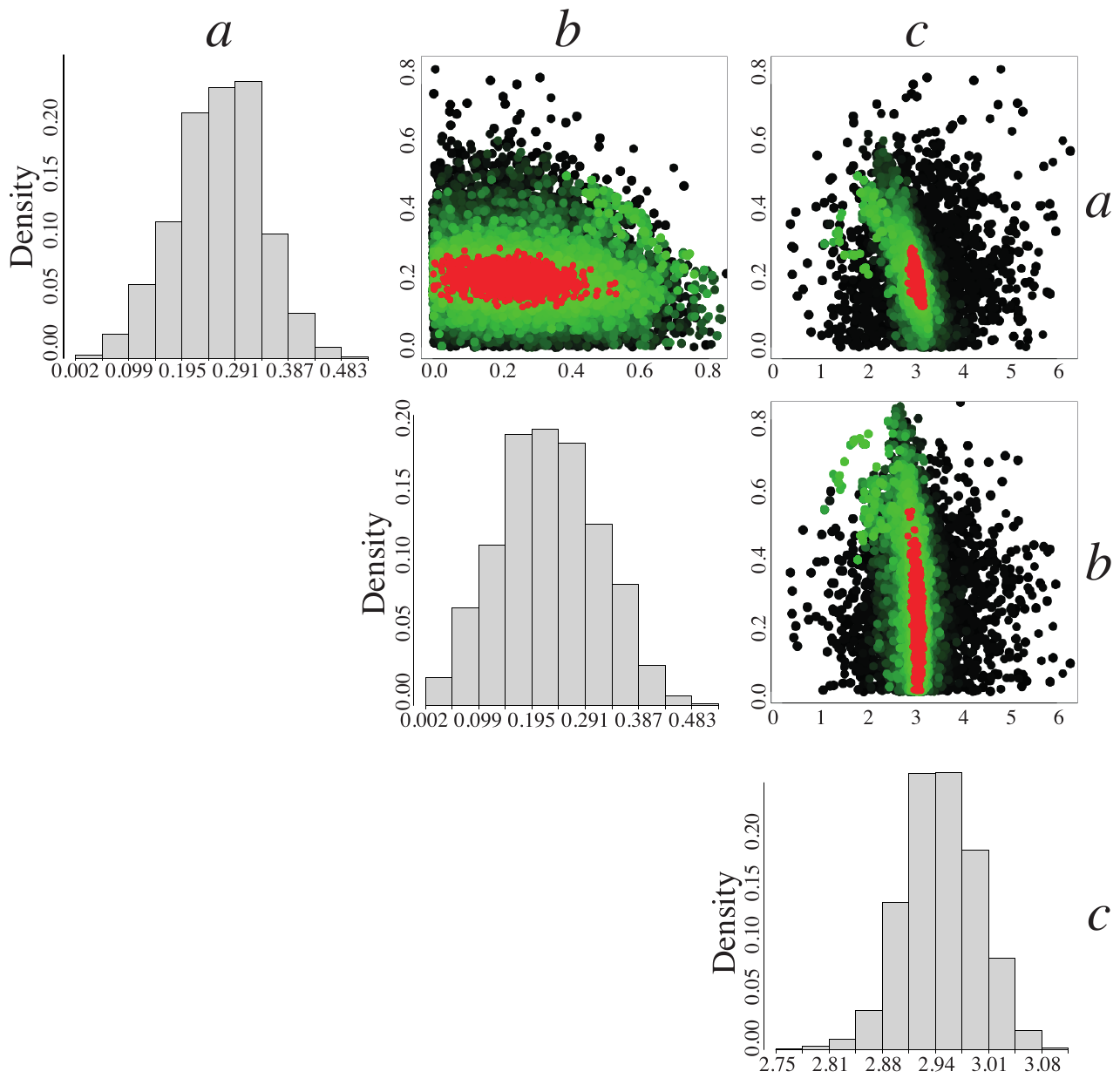}}
\caption[Fitzhugh-Nagumo model. Scatter plots and marginal distributions]{2D scatterplots and marginal distributions of the three parameters $a,b,c$ of the Fitzhugh-Nagumo model over 50 populations. The initial particles, coloured black in the scatter plots, are sampled from a component-wise normal prior with a positive truncation. Particles of the intermediate populations are represented in increasingly brighter shades of green with particles of the final population, representing the posterior, coloured red.}
\label{fnscatter}
\end{figure}

\begin{figure}[h] 
\centering
{\includegraphics{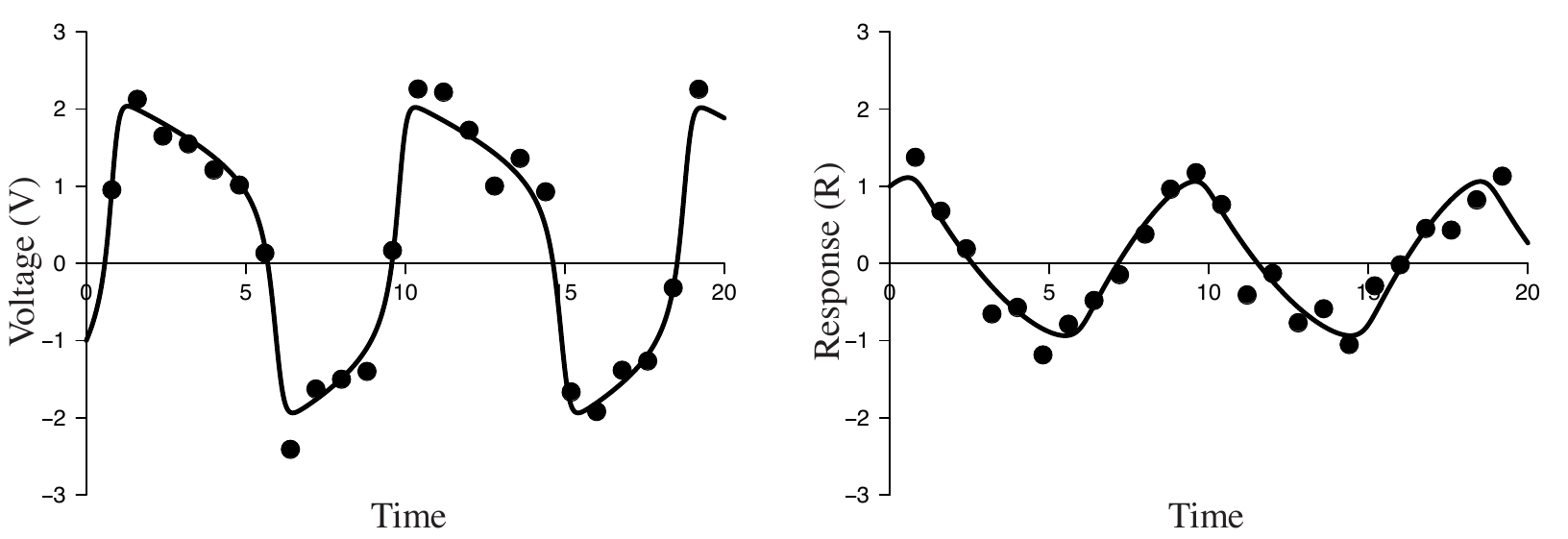}}
\caption[Fitzhugh-Nagumo model Voltage/Response time series]{Voltage ($V$) and Response ($R$) observations (dots) and inferred time-series curves (lines). The expected time-series curve is obtained by taking the weighted mean of the inferred curves across all particles from the final population.}
\label{fnts}
\end{figure}

Using $N=1000$ particles, a discretisation step-size $\epsilon=0.6$, and a resampling threshold of $0.3$, we perform the algorithm over 50 tempered distributions \eqref{temperedds} in a geometric sequence \eqref{geometric}. We adopt component-wise normal priors for the three parameters. The prior is centred on the simulated mode, i.e. $(\mu_a, \mu_b, \mu_c) = (0.2,0.2,3)$, and we set the covariance matrix to be $\Sigma=\diag(0.3^2, 0.3^2, 1.5^2)$. In addition we impose a positivity constraint on the parameters.  The derivatives of $\dot{V}$ and $\dot{R}$, required for the calculation of the Fisher metric, are given in \ref{pdode}. We present the scatter plots and weighted histograms representing the full and marginal inferred posterior in Figure \ref{fnscatter} and plot the resulting expected ODE solutions in Figure \ref{fnts}. The results verifies the applicability of the SMC sampler to dynamical systems parameter inference.

\begin{figure}[h] 
\centering
{\includegraphics{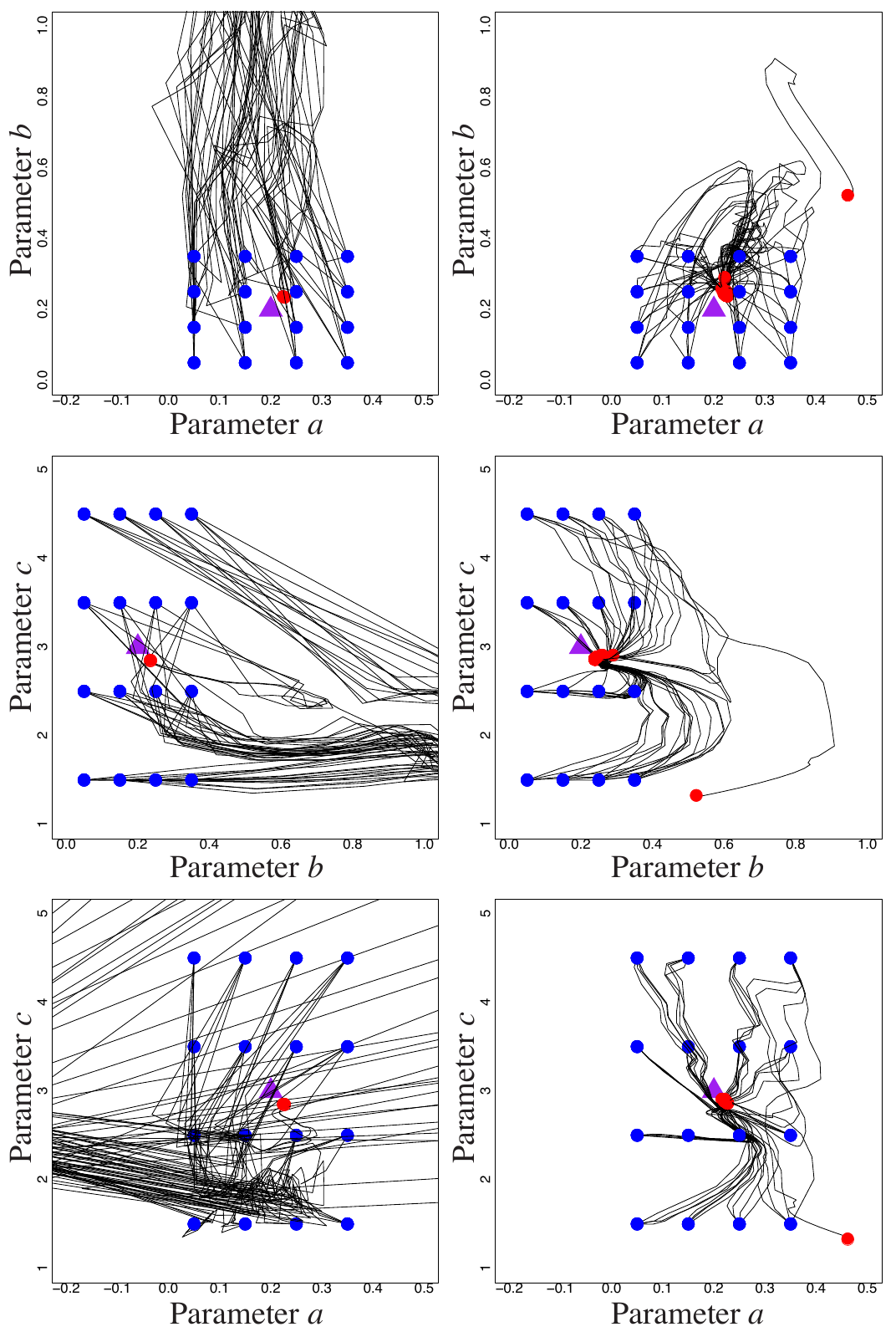}}
\caption[Fitzhugh-Nagumo model. mMALA paths]{mMALA particle drift paths for the Fitzhugh-Nagumo model. Each of the three rows represent a projection to a separate two-dimensional parameter subspace; the left and right columns are simulations with a uniform and a normal prior respectively. The lack of a lower bound on the Fisher metric in the simulation with the uniform prior (see text) results in a drift behaviour consistent with an extremely high temperature regime, rendering the SMC sampler algorithm unworkable.}
\label{fndrift}
\end{figure}

We turn now to the issue of the identity of the prior distribution, the importance of which we have already alluded to at various points in our discussion (c.f. \eqref{unimetric} and Section \ref{uniparam}). Similar to the simulations of the static Gaussian example in the previous section, we select an arbitrary grid of points about the simulation mode and focus on the deterministic drift by ignoring the diffusion steps and accepting all particles. We compare the behaviour arising from the component-wise normal prior given above and with that from a component-wise uniform prior with lower and upper limits $(a_l, b_l, c_l)=(0,0,0)$ and $(a_u, b_u, c_u)=(1,1,7)$ respectively. The plots of the particle drifts are given in Figure \ref{fndrift}. It is clear that, unlike the case with the normal prior, the SMC sampler with a uniform prior does not lead to an orderly drift of particles toward the simulated mode, which results in a highly inefficient sampler where the particle acceptance rate at each MH-step is extremely low ($< 1\%$). This occurrence is unfortunate, if not entirely unexpected for the following reason. The procedure of utilising tempered distributions implies that the components of the Fisher metric \eqref{gs} corresponding to the initial distributions with low values of the tempering parameter $\phi_a$ are bounded from below by a function of the prior parameters, i.e. for small $\phi_a$,
\begin{equation}
(g_a)_{ij} \approx \expt_{\xi}(h_{ij}).
\end{equation}
For a uniform prior, $h_{ij}=0$ and hence $(g_a)_{ij} \approx 0$ for all $i,j$. Given that both the drift and diffusion terms (\eqref{proposal}, \eqref{mu}) are proportional to the inverse metric $g_a^{-1}$, the observed high-temperature drift behaviour\footnote{$\phi_a\propto \frac{1}{kT}$, where $T$ is the temperature.} in Figure \ref{fndrift} is fully consistent with the theoretical expectations.

In our simulations, it turns out that the SMC sampler with mMALA transitions is not much more robust than one with a non-differential geometric global adaptive kernel, the latter with the computational benefit of not requiring costly computations of the Fisher metric. We speculate on the possible reasons in the next section. However, we now consider another example where the performance of the information geometric SMC sampler shows a clear advantage over a standard adaptive kernel.

\paragraph{Example: Lotka-Volterra Model}
The Lotka-Volterra model is a simple representation of the predator-prey relationship in a closed environment. Let the variables $x, y$ represent the numbers of prey and predator species respectively. The dynamics are represented by the set of ODEs
\begin{align}
\frac{\dd x}{\dd t} &= x(\alpha - \beta y),\\
\frac{\dd y}{\dd t} &= -y(\gamma - \delta x),
\end{align} 
where the four parameters $(\alpha, \beta, \gamma, \delta)$ represent the prey birth rate, prey death rate due to predation, the predator death rate, and the predation efficiency respectively.

\begin{figure}[h] 
\centering
{\includegraphics{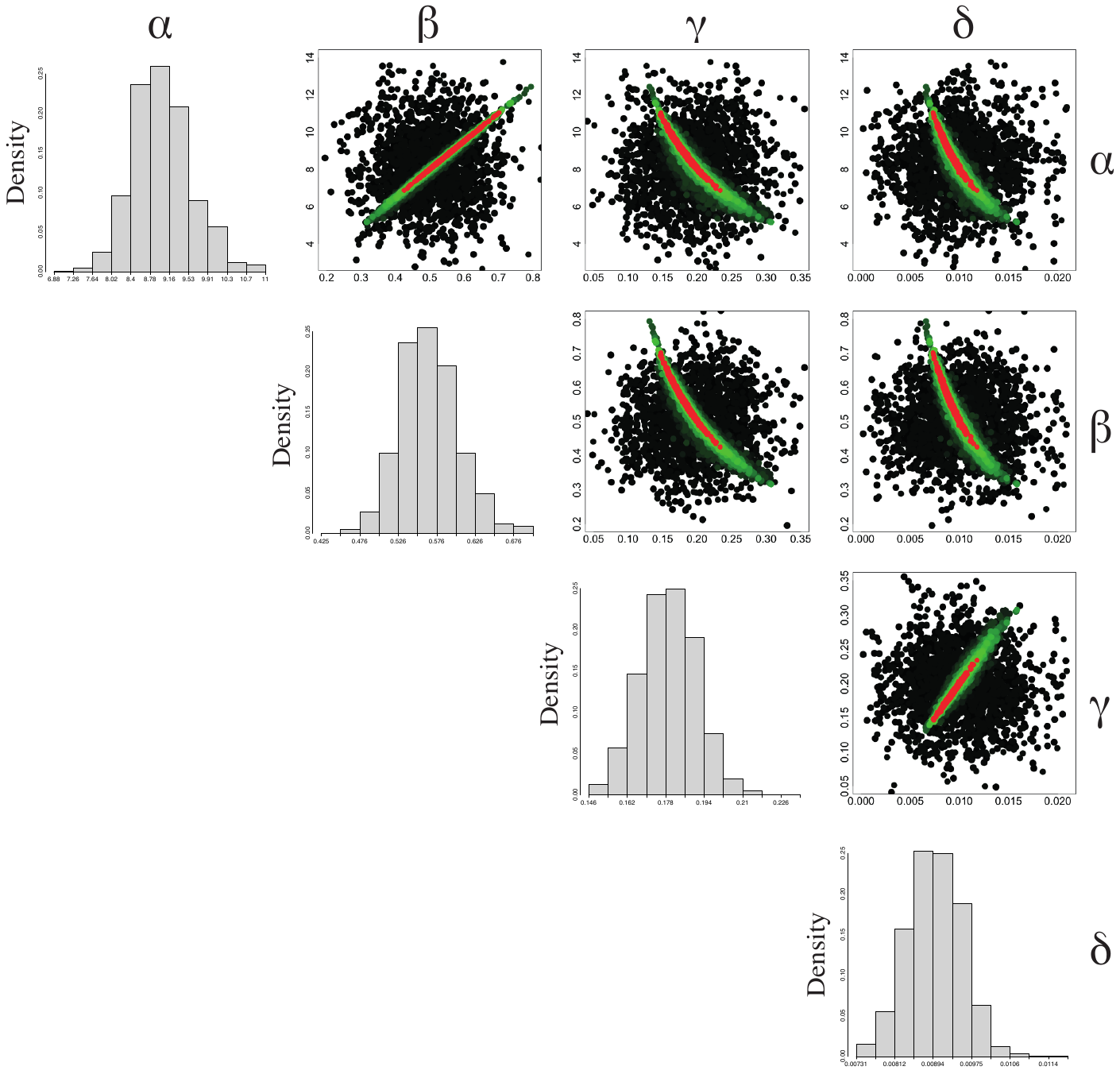}}
\caption[Lotka-Volterra model. Scatter plots and marginal distributions]{2D scatterplots and marginal distributions of the four parameters $\alpha, \beta, \gamma, \delta$ of the Lotka-Volterra model over 30 populations. The initial particles are sampled from a component-wise uniform prior with limits as shown in the figures. These particles are coloured black in the scatter plot. Particles of the intermediate populations are represented in increasingly brighter shades of green with particles of the final population, representing the posterior, coloured red.}
\label{lvscatter}
\end{figure}

\begin{figure}[h] 
\centering
{\includegraphics{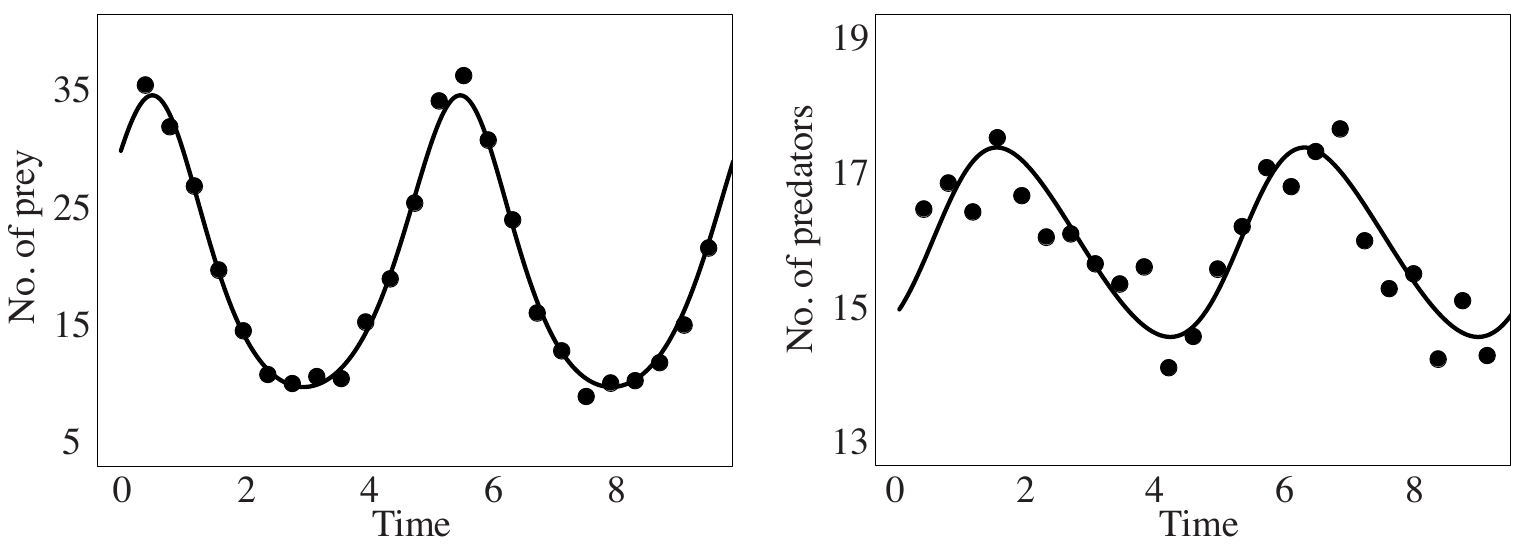}}
\caption[Lotka-Volterra model predator/prey time series]{Predator and prey observations (dots) and inferred time-series curves (lines). The expected time-series curves are obtained by taking the weighted mean of the inferred curves across all particles.}
\label{lvts}
\end{figure}

We set up a simulation with starting population $(x_0, y_0)=(15, 30)$ and parameters $(\alpha', \beta', \gamma', \delta') = ( 8, 0.5, 0.2, 0.01)$. We adopt the same assumptions as for the Fitzhugh-Nagumo model simulation but with the normal noise $\sigma^2=0.4$, and prior parameters centred on the simulated modes with $\Sigma=\diag(2^2, 0.1^2, 0.05^2,0.004^2)$. We let the number of particles $N= 1000$, the discretisation step-size $\epsilon = 0.5$, and a resampling threshold of $0.3$. We perform the algorithm over 30 tempered distributions \eqref{temperedds} in a geometric sequence \eqref{geometric}. In Figure \ref{lvscatter}, we show the scatter plots and marginal weighted histograms of the inferred posterior. It is clear from the scatter plots that the parameters are very highly correlated -- precisely the regime where the information geometric approach is expected to confer the greatest benefit. We verify the accuracy of the sampler by simulating the curves with the particle parameters, as shown in Figure \ref{lvts}.

To demonstrate the efficiency of the differential geometric SMC kernel we benchmark its performance with a non-differential geometric, but nevertheless adaptive, kernel. We consider an MCMC kernel with an MVN proposal $K'(\xi_{a+1}\mid \xi_{a})$ with zero mean and a covariance matrix set to the sample covariance matrix of the particles in population $a$ multiplied by the asymptotic factor $(2.38)^2/D$ \cite{Atchade:2011wf}. Here $D$ is the dimension of the parameter space (i.e. $D=4$). 

\begin{figure}[h] 
\centering
{\includegraphics{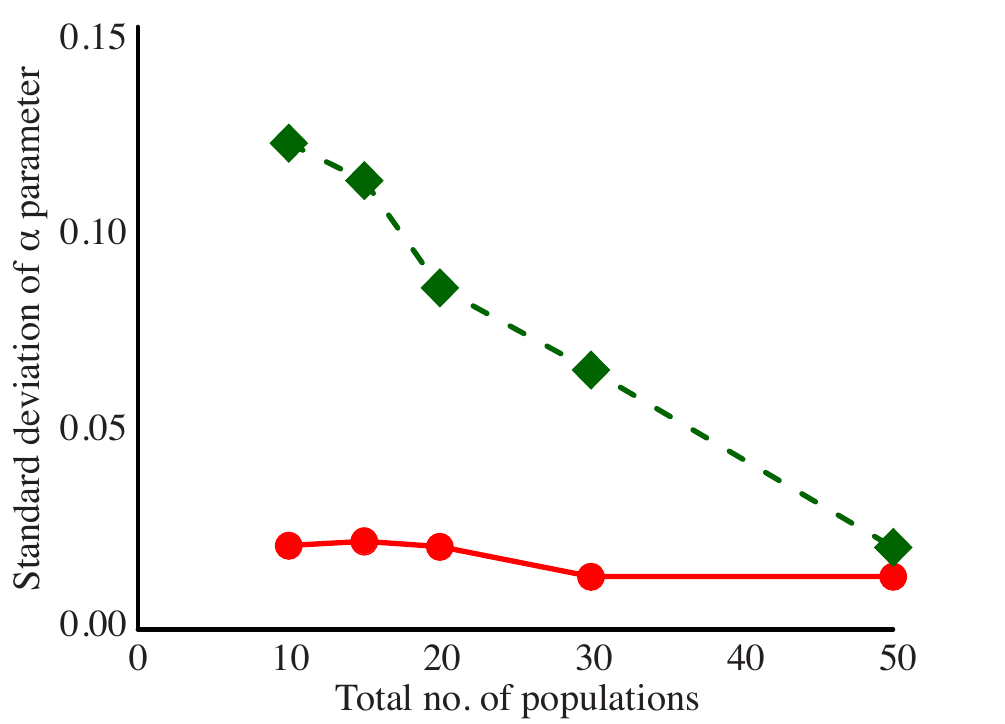}}
\caption[Variance in posterior expectation for $\alpha$]{The variance in the estimate of the $\alpha$ parameter of the Lotka-Volterra model as given by SMC performed using mMALA proposals (\textbf{red} $\bullet$) and non-differential geometric global adaptive kernel (\textbf{green} $\blacklozenge$). The profile of the estimates of the other three parameters $(\beta, \gamma, \delta)$ are similar.} 
\label{poscomp}
\end{figure}

The one measure of efficiency of practical importance we would like to examine is robustness; in the context of parameter inference this means a low variability in the inferred statistics over repeated runs of the SMC sampler. To that end we run, for both the mMALA and the above benchmark kernel,  the SMC sampler for a range of total intermediate populations, performing 27 repetitions of each run. Over each set of 27 runs we determine the sample variance of the inferred parameter means. For both kernels, one would expect the sample variance to decrease with increasing total number population. This is, indeed, what we observe, and the results for parameter $\alpha$ are given in Figure \ref{poscomp}. We see that the mMALA kernel outperforms the non-differential geometric adaptive kernel by demonstrating a consistently high level of robustness over different number of intermediate distributions. By employing the mMALA kernel, we have effectively shortened the distance between successive distributions in the SMC chain, thereby replicating the algorithm with a greater number of intermediate distributions.

\begin{figure}[h] 
\centering
{\includegraphics{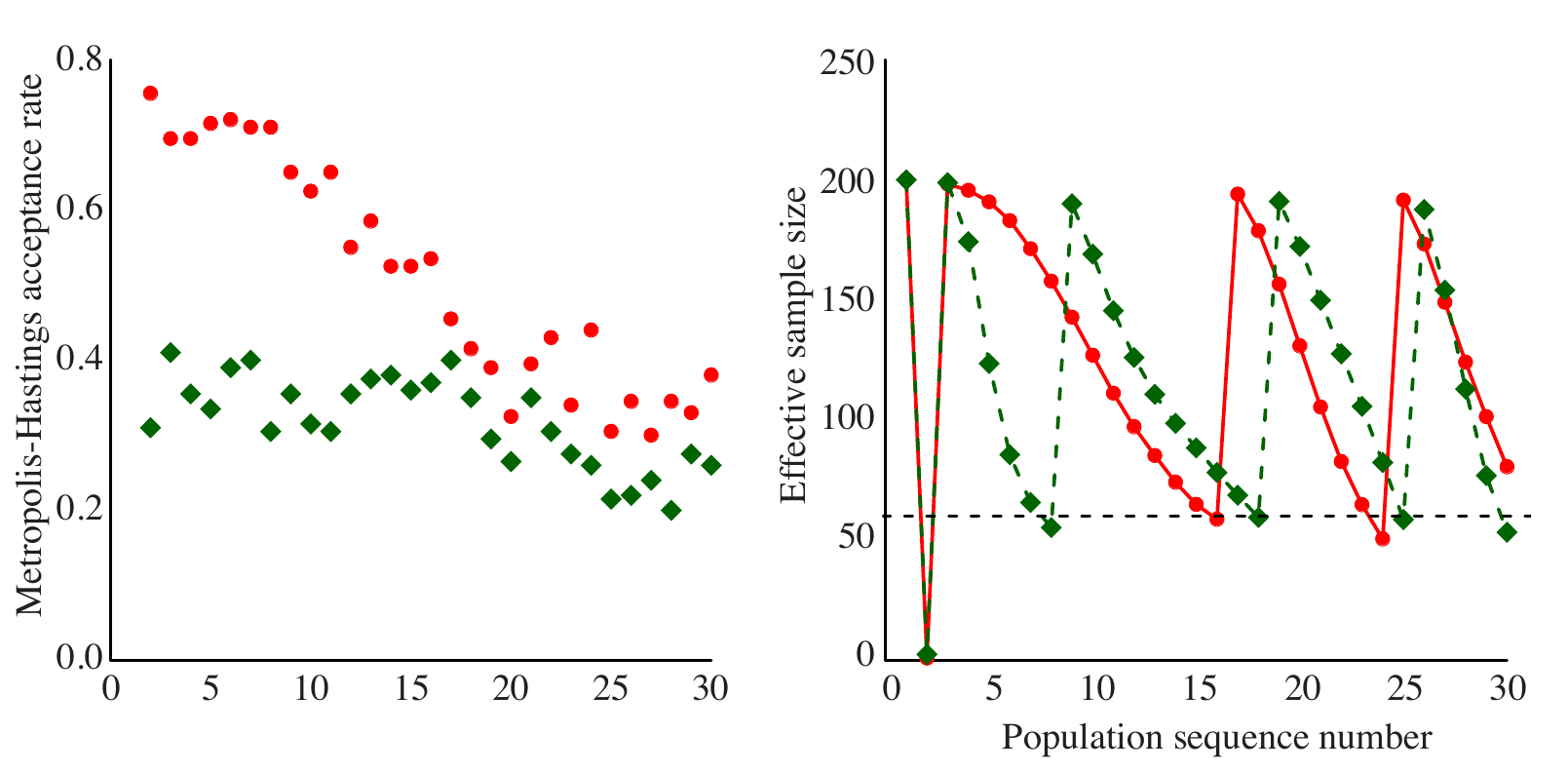}}
\caption[ESS and acceptance rates]{The Metropolis-Hastings acceptance rates (left) and effective sample sizes (right) over a representative single run of the SMC sampler employing the mMALA kernel  (red $\bullet$) and the non-differential geometric adaptive random walk kernel (green $\blacklozenge$). The horizontal dotted line indicates the 30\% resampling threshold. Employing the mMALA kernel results in a higher MH-acceptance rate and a correspondingly more gradual decline in the ESS.}
\label{effsamplv}
\end{figure}

An alternative, explanatory, view of the efficiency of the mMALA kernel can be seen by tracking the effective sample sizes and particle acceptance rates of the SMC sampler over the intermediate populations. This is shown in Figure \ref{effsamplv}. The relatively gradual decline in the ESS of the SMC populations when the mMALA kernel is employed is similar to the observed behaviour ESS along the geodesic in Figure \ref{geodesicpath}.

\section{Summary and Discussion}\label{discussion}
In this paper we have explored the application of methods and ideas from information geometry to SMC sampling and have demonstrated its use in sequential Bayesian inference. We focused on two areas -- the construction of the sequence of distributions which lie on geodesics, and the employment of mMALA as MCMC transition kernels between the intermediate distributions.

Of the two areas, the theoretical foundations of the mMALA is more established and we have shown that the extension from MCMC to SMC is relatively straightforward. In particular the theoretical differences are the replacement of the single Fisher metric with a sequence of Fisher metrics defined by the sequence of distributions, and the parallelisation of the mMALA diffusion over multiple particles. The conceptual similarity with its use in MCMC allows all the advantages of the original mMALA formulation in Girolami and Calderhead \cite{Girolami:2011hw} -- namely the self-tuning characteristics, efficiency in highly-correlated and high-dimensional settings -- to be carried across to SMC. 

The issue of intermediate distributions on geodesics is, in comparison, more difficult to tackle, simply due to the lack of analytical solutions for most non-trivial examples. Nevertheless we have demonstrated the potential advantages, in the form of a slower decline in the effective sample sizes, of placing our SMC intermediate distributions on geodesics by way of a simple univariate normal inference example. The computational price of evaluating these geodesics (where available) is a one-off cost incurred at the start of the SMC run and is independent of the number of sampled particles.

At present the computational overhead involved in determining the components of the Fisher metric is seemingly a disadvantage. In the application to the analysis of dynamical systems, one is required to solve at least two separate sets of differential equations for the sensitivities and their derivatives; for heteroscedastic observation noise models, the burden is compounded by the need to solve for the double derivative of the metric. Because the computational cost of evaluating the Fisher metric scales as $\Ord(D^2)$, with $D$ the dimension of the statistical manifold, the efficiency of the technique is reduced in precisely the high-dimensional regime that a non-information geometric kernel might be less suited. For more complicated noise models (e.g. heteroscedasticity), the computational complexity scales as $\Ord(D^3)$ -- this is a consequence of the need to calculate the double derivatives $\partial_i\partial_jS_k$ (see \ref{mmaladev}). The computational burden can be reduced by adopting the simplified mMALA, where only the first-order sensitivities are calculated \cite{Girolami:2011hw, BenCalderhead:2011vx}. However, as shown in the original MCMC context \cite{Girolami:2011hw} the effective computational time of the simplified mMALA, expressed in units of the ESS, is not significantly better than its full counterpart.  Nevertheless, even in low-dimensional problems, especially in situations where the parameters are highly correlated, as we observed in the Lotka-Volterra model, the advantages conferred by the information geometric approach are clearly evident. Furthermore, it has been suggested\footnote{By A. C. C. Coolen in the Discussion section of \cite{Girolami:2011hw}.} that the calculation of the Fisher metric can be made more efficient via a canonical transformation of the coordinates. This is an interesting avenue for further research.

When employing the SMC sampler in sequential Bayesian inference, the form of the prior is often chosen primarily on the basis of knowledge of the parameter constraints, for example a gamma distribution for strictly positive parameters, or a uniform prior when knowledge of the upper and lower bounds is available. Although such bounded priors enforce weak identifiability on the parameter estimation, as was highlighted in \cite{BenCalderhead:2011vx} and \cite{Vallisneri:2007cx}, the prior now has the additional role of regularising the Fisher metric. Indeed, as we have shown via the Fitzhugh-Nagumo model example, choosing a uniform prior in conjunction with a tempered sequence of distributions severely impacts the viability of the mMALA kernels; flat or nearly-flat intermediate distributions, as is the case with the initial distributions, results in high-temperature diffusion processes. Because the mMALA diffusion process is informed by the local geometry of the statistical manifold, it does not respect the global density boundaries and attempts to scatter the particles to fill the entire untruncated support, resulting in very low MH-acceptance rates.

Although the transition kernels adapt to the local geometry of the manifold, there is still the outstanding issue of tuning the discretisation step-size parameter. We have shown in the univariate model example that there is a minimum step-size, below which the particles do not travel far enough over the course of the SMC run. One possibility for improvement might be to adopt a higher-order discretisation approximation of the Langevin diffusion SDE which will allow us to select larger step size $\epsilon$ without deviating too significantly from the Langevin diffusion path, which, in turn, leads to the maintenance of a high MH-acceptance rate and a reduction in the number of intermediate distributions needed for a given level of robustness. This can be achieved by replacing the first-order Euler discretisation with the higher-order Ozaki discretisation \cite{Ozaki:1992tw}. We provide a brief description of the Ozaki discretisation in \ref{Oz}. At present it is not clear if the advantages are negated by the added complexity and the resulting increased computational burden, or if the performance of the simplified mMALA relative to the full version can be improved by altering the discretisation scheme.

In summary, we see that the methods of information geometry can be applied to SMC samplers allowing for adaptive and efficient transition kernels. The information theoretic formulation provides a neat and aesthetically pleasing geometrical framework for future improvements in the algorithm. However we have shown that mMALA kernels cannot substitute for careful monitoring of convergence, and potentially adjusting kernels appropriately. This is particularly challenging in areas where likelihoods are flat, which includes many dynamical systems \cite{Gutenkunst:2007aa,Apgar:2010di,Erguler:2011bu}.

\appendix
\section{Derivations for the geodesic example}
\subsection{Geodesics on the Gaussian distribution manifold}\label{geoderive}
Consider a multivariate normal distribution $(\mu, \Sigma)$ defining a manifold $\mathcal{S}$. Following the analysis in \cite{Calvo:1991tf, Skovgaard:1984uk, Imai:2011wo}, the metric on this space can be written using \eqref{fisheralt} as 
\begin{equation}
\dd s^2 = (\dd\mu)^T\Sigma^{-1}\dd\mu + \frac{1}{2}\tr(\Sigma^{-1}\dd\Sigma)^2.
\end{equation}
In these coordinates, the geodesic equations \eqref{geodesiceom} are written as 
\begin{align} \label{orinormgeo1}
\ddot{\Sigma} + \dot{\mu}^T\mu - \dot{\Sigma}\Sigma^{-1}\dot{\Sigma} &= 0,\\
\ddot{\mu} - \dot{\Sigma}\Sigma^{-1}\dot{\mu} &= 0.\label{orinormgeo2}
\end{align}
The strategy is to first solve for geodesics through the origin $(\mu(0), \Sigma(0)) = (0, \mathbb{I}_p)$ before translating to the curve with the desired end-points. Adopting the canonical coordinates $(\Delta, \delta) \equiv (\Sigma^{-1}, \Sigma^{-1}\mu)$, \eqref{orinormgeo1} and \eqref{orinormgeo2} can be partially integrated to give
\begin{align}
\dot{\Delta} &= -B\Delta + x\delta^T,\\
\dot{\delta} &= -B\delta + (1+ \delta^T\Delta^{-1}\delta)x,
\end{align}
where $B=\dot{\Delta}(0)$ and $x=\dot{\delta}(0)$, and $\Delta(t)$ and $\delta(t)$ are the geodesic coordinates parametrised by  $t\in\mathbb{R}$. It can be shown \cite{Imai:2011wo} that the solutions to the geodesic equations are
\begin{align}
\begin{split}
\Delta(t) &= \mathbb{I}_p +\frac{1}{2}\bigl[\cosh(tG) - \mathbb{I}_p\bigr] + \frac{1}{2}B\bigl[\cosh(tG) - \mathbb{I}_p\bigr](G^{-1})^2B\\
& \qquad\qquad- \frac{1}{2}\sinh(tG)G^{-1}B - \frac{1}{2}B\sinh(tG)G^{-1},
\end{split}\\
\delta(t) &= -B\bigl[\cosh(tG) - \mathbb{I}_p\bigr](G^{-1})^2x + \sinh(tG)(G^{-1}x),
\end{align}
where $G^2 := B^2 + 2xx^T$. Simplifying to one dimension $(p=1)$, we have
\begin{align}
\Delta(t) &= 1+ \frac{1}{2}(1+R^2)\bigl(\cosh(tG)-1\bigr) - R\sinh(tG),\label{A8}\\
\delta(t) &= \Bigl(\frac{1-R^2}{2}\Bigr)^{\frac{1}{2}}\bigl(-R(\cosh(tG)-1) +\sinh(tG)\bigr),\label{A9}
\end{align}
where $R=B/G$. This solution would suffice, except that it is given in terms of the gradient terms $B, G$ instead of the target end-point coordinates; it is however not difficult to rewrite the solution. Keeping $(p=1)$, we set, without loss of generality, the target end-point of the geodesic to be located at $t=1$. Let $\Delta'\equiv \Delta(1)$ and $\delta' = \delta(1)$ we solve \eqref{A8} and \eqref{A9} for $R$ and $G$ giving
\begin{align}
R &= \frac{\delta'^2 - 2\Delta'^2 + 2\Delta}{(\delta'^4 + 4\delta'^2\Delta'^2 + 4\delta'^2\Delta' + 4\Delta'^4 -8\Delta'^3 + 4\Delta'^2)^\frac{1}{2}},\label{A10}\\
G &= \cosh^{-1}\Bigl(\frac{\delta'^4 + 4\delta'^2\Delta'^2 + 4\delta'^2\Delta' + 4\Delta'^4 + 4\Delta'^2}{8\Delta'^3}\Bigr).\label{A11}
\end{align}
Substituting these expressions for $R$ and $G$ back into \eqref{A8} and \eqref{A9} gives us the solution for geodesics through the origin and a specified point coordinate at $t=1$.

Now in order to translate geodesics to geodesics, we need a map $g: \mathcal{S} \rightarrow \mathcal{S}$ which leaves the Fisher metric invariant. As shown in \cite{Imai:2011wo} the map which achieves this is the symmetric group $GA^+(p)/SO(p)$ where the positive affine group is defined as
\begin{equation}
GA^+(p) := \{ g=(d,P) \in\mathbb{R}^p \times GL(p, \mathbb{R}) | \det P >0\},
\end{equation}
and the group action on points in $\mathcal{S}$, in terms of coordinates $(\mu, \Sigma)$ and $(\Delta, \delta)$, are 
\begin{equation}
\begin{split}
\mathcal{S}\quad&\longrightarrow\quad\mathcal{S}\\
(\mu, \Sigma)\quad&\longmapsto\quad(P\mu + d, P\Sigma P^T)\\
(\Delta, \delta)\quad&\longmapsto\quad\bigl((P^{-1})^T\Delta P^{-1}, (P^{-1})^T\Delta P^{-1}\Delta^{-1}\delta + \Delta P\Delta^{-1}\delta + \Delta d\bigr),
\end{split}
\end{equation}
and with inverse element $g^{-1} = (-P^{-1}d, P^{-1})$.

Given two points $p_1, p_2\in \mathcal{S}$, it is then straightforward to solve for the desired geodesic. First, obtain the group element $g'\in GA^+(p)/SO(p)$ which maps the origin to $p_1$. Substituting $(\Delta', \delta') = g'^{-1}p_2$ into \eqref{A8} and \eqref{A9} via \eqref{A10} and \eqref{A11}, we obtain a solution through the origin which can then be translated using $g'$ to describe the desired geodesic.

\subsection{Kernel density for a uniform proposal and Gaussian target}\label{kernelderive}
The kernel density in \eqref{speker} is the Metropolis-Hastings algorithm acceptance probability \cite{Metropolis:1953in} for a uniform proposal of width $d$ and a univariate Gaussian target. Following the MH-algorithm, the expression is calculated separately for the cases where the proposal is accepted or rejected. We have
\begin{equation}
K_a(\xi_{a}\, |\, \xi_{a-1}) =
\begin{cases}
 1 - \max\Bigl[0,\, \frac{1}{d\gamma_a(\xi_{a-1})}\Bigl(\Phi\Bigl(-\frac{|\mu - \xi_{a-1}|}{\sigma}\Bigr)- \Phi\Bigl(-\frac{\xi_{a-1} - \mu - a/2}{\sigma}\Bigr) \Bigr] \\
\quad- \max\Bigl[0,\, \frac{1}{d\gamma_a(\xi_{a-1})}\Bigl(\Phi\Bigl(-\frac{\xi_{a-1} - \mu + a/2}{\sigma}\Bigr)- \Phi\Bigl(-\frac{|\mu-\xi_{a-1}|}{\sigma}\Bigr) \Bigr] \\
 \quad- \min\Bigl[\frac{1}{2}, \,\frac{2}{d}|\mu - \xi_{a-1}|\Bigr],  
&(\text{for } \xi_a = \xi_{a-1})\\ \\

\frac{1}{d}\min\Bigl[1, \, \exp\Bigl(-\frac{(\xi_a - \mu)^2}{2\sigma^2} + \frac{(\xi_{a-1} - \mu)^2}{2\sigma^2}\Bigr)\Bigr], & (\text{for } \xi_a \neq \xi_{a-1})
\end{cases}
\end{equation}
with $(\mu, \sigma)$ the mean and standard deviation parameters of the $a$th intermediate distribution. $\xi_a$ is the particle coordinate of population $a$ and $\Phi$ is the cumulative distribution function of the standard normal.

\section{Non-linear ODE calculations}\label{nonlinearode}

\subsection{Auxiliary differential equations for the sensitivities $S_{i, n}$}\label{sensitivityDE}
The subscript conventions used in this section indicates the variable in the differential. We have $\{i, j, k\}\sim \xi$ and $\{l, m, n\}\sim X$, for example $\partial_j\partial_lf \equiv \frac{\partial^2f}{\partial\xi^j\partial X^l}$. To avoid cluttering the notation, when there is no ambiguity, we drop the $X$-index on $\dot{S}$, i.e. $\dot{S}_{i,l}\rightarrow \dot{S}_i$ and $f_l\rightarrow f$. 
\begin{align}
\dot{S}_i &= (\partial_l f)S_{i,l} + \partial_i f,\label{sdot}\\
\partial_k\dot{S}_i &= (\partial_m\partial_lf) S_{k,m}S_{i,l} + (\partial_k\partial_l f)S_{i,l} + (\partial_l f)(\partial_kS_{i,l}) + (\partial_l\partial_if)S_{k,l} + \partial_i\partial_kf,\label{dsdot}\\
\begin{split}
\partial_j\partial_k\dot{S}_i &= (\partial_m\partial_l\partial_j f)S_{k,m}S_{i,l} + (\partial_m\partial_l f)(\partial_j S_{k,m})S_{i,l} + (\partial_m\partial_lf)(\partial_j S_{i,l})S_{k,m}\\
 &\quad+ (\partial_k\partial_j\partial_lf)S_{i,l} + (\partial_k\partial_lf)(\partial_j S_{i,l}) + (\partial_l\partial_j f)(\partial_k S_{i,l}) + (\partial_l f)(\partial_j\partial_k S_{i,l})\\
 &\quad + (\partial_l\partial_j\partial_i f)S_{k,l} + (\partial_l\partial_i f)(\partial_j S_{k,l}) + \partial_i\partial_j\partial_kf\\
 &\quad + S_{j,n}\bigl[(\partial_m\partial_l\partial_nf)S_{k,m}S_{i,l} +  (\partial_k\partial_l\partial_n f)S_{i,l} + (\partial_l\partial_n f)(\partial_k S_{i,l})\\
 &\qquad\qquad+ (\partial_l\partial_i\partial_nf) S_{k,l} + \partial_i\partial_k\partial_n f\bigr].
\end{split}\label{ddsdot}
\end{align}

\subsection{Partial derivatives for the ODE examples}\label{pdode}
\paragraph{Fitzhugh-Nagumo model}
We reproduce here the expressions in \cite{Girolami:2011hw}. The two components are labelled $(V, R)$ with parameters $(a,b,c)$. The non-zero single derivatives are
\begin{equation}
\begin{split}
\frac{\partial\dot{V}}{\partial c} &= V - \frac{V^3}{3} + R, \quad \frac{\partial\dot{R}}{\partial a} = \frac{1}{c}, \quad \frac{\partial\dot{R}}{\partial b} = -\frac{R}{c}, \quad \frac{\partial\dot{R}}{\partial c} = \frac{V -a +bR}{c^2},\\
\frac{\partial\dot{V}}{\partial V} &= c(1-V^2),\quad \frac{\partial\dot{V}}{\partial R} = c,\quad \frac{\partial\dot{R}}{\partial V} = -\frac{1}{c},\quad \frac{\partial\dot{R}}{\partial R} = -\frac{b}{c},  
\end{split}
\end{equation}
and the double derivatives 
\begin{equation}
\begin{split}
\frac{\partial^2\dot{R}}{\partial^2c} &=\frac{2(a-V -bR)}{c^3},\quad \frac{\partial^2\dot{R}}{\partial a\partial c} = -\frac{1}{c^2}, \quad \frac{\partial^2\dot{R}}{\partial b\partial c} = \frac{R}{c^2},\\
\frac{\partial^2\dot{V}}{\partial V\partial c} &= 1-V^2, \quad \frac{\partial^2\dot{V}}{\partial R\partial c} = 1,\quad \frac{\partial^2\dot{R}}{\partial R\partial b} = -\frac{1}{c},\quad \frac{\partial^2\dot{R}}{\partial R\partial c} = \frac{b}{c^2}.
\end{split}
\end{equation}

\paragraph{Lotka-Volterra model}
Let $(x, y)$ label the prey and predator respectively with parameters $(\alpha, \beta, \gamma, \delta)$. The non-zero single derivatives are
\begin{equation}
\begin{split}
\frac{\partial\dot{x}}{\partial\alpha} & = x, \quad \frac{\partial\dot{x}}{\partial\beta} = -xy, \quad \frac{\partial\dot{y}}{\partial\gamma} = -y, \quad \frac{\partial\dot{y}}{\partial\delta}  = xy,\\
\frac{\partial\dot{x}}{\partial x}& = \alpha -\beta y, \quad \frac{\partial\dot{x}}{\partial y} = -\beta x, \quad \frac{\partial\dot{y}}{\partial x}  = \delta y, \quad \frac{\partial\dot{y}}{\partial y}  = -(\gamma- \delta x), \\
\end{split}
\end{equation}
and the double derivatives are
\begin{equation}
\begin{split}
\frac{\partial^2 \dot{x}}{\partial x\partial\alpha} = -\frac{\partial^2 \dot{y}}{\partial y\partial\gamma} = 1, \quad \frac{\partial^2 \dot{x}}{\partial x\partial\beta} &= -\frac{\partial^2 \dot{y}}{\partial x\partial\delta} = -y, \quad \frac{\partial^2 \dot{x}}{\partial y\partial\beta} = -\frac{\partial^2 \dot{y}}{\partial y\partial\delta} = -x,\\
\frac{\partial^2 \dot{x}}{\partial x\partial y} &= -\beta, \quad \frac{\partial^2 \dot{y}}{\partial x\partial y} = \delta.
\end{split}
\end{equation}

\section{Extensions of mMALA}

\subsection{Heteroscedasticity and bounded parameters}\label{mmaladev}
In this section we present the implications on the calculation of the Fisher metric of incorporating additional structure on top of the normal noise model. In particular, we focus on heteroscedasticity and boundary constraints in observation noise. In the first case, in addition to a fixed normal observation noise, we have a noise contribution which scale scales with $X$; here the noise model is given by \eqref{normL} but with the replacement
\begin{equation}
\Sigma_d :=   \mathbb{I}_\tau\sigma_d^2 + \diag\bigl((\mathbb{I}_\tau \tilde{\sigma}_d^2)(X_{\cdot, d} *X_{\cdot, d})\bigr).\label{sigmahet}
\end{equation}
In the second case, because $Y$ often represents measurement of physical quantities (concentration, volume, etc) there is usually a positive constraint on $X$ and $Y$, the implication being that one has to consider truncated distributions.

Using the expression of $\Phi$ from \eqref{normL} for the multivariate normal and a simple application of the chain rule, we have
\begin{equation}
\begin{split}
-\partial_i\partial_j\Phi_d & = \partial_i\partial_j(\frac{1}{2}(Y_d-X_d)^T\Sigma^{-1}_d(Y_d-X_d))\\
&=S_{d,i}^T\Sigma^{-1}_dS_{d,j} - (Y_d-X_d)^T\bigl(\Sigma^{-1}_d(\partial_iS_{d,j}) + (\partial_i\Sigma^{-1}_d)S_{d,j} + (\partial_j\Sigma^{-1}_d)S_{d,i}\bigr)\\
& \quad+ \frac{1}{2}(Y_d-X_d)^T(\partial_i\partial_j\Sigma^{-1}_d)(Y_d-X_d).
\end{split}
\end{equation}
For a normal distribution truncated between limits $a < X < b$, the expectation and variance can be written in terms of the untruncated mean and s.d. $(\mu, \sigma)$ as \cite{Johnson:1995tp}
\begin{align}\label{B1}
\E(X | a <X<b) &= \mu +\frac{\phi(\frac{a-\mu}{\sigma}) - \phi(\frac{b-\mu}{\sigma})}{\Phi(\frac{b-\mu}{\sigma}) - \Phi(\frac{a-\mu}{\sigma})}\sigma,\\
\Var(X | a<X<b) &= \sigma^2\biggl[1+ \frac{\frac{a-\mu}{\sigma}\phi(\frac{a-\mu}{\sigma})-\frac{b-\mu}{\sigma}\phi(\frac{a-\mu}{\sigma})}{\Phi(\frac{b-\mu}{\sigma})-\Phi(\frac{a-\mu}{\sigma})} - \biggl(\frac{\phi(\frac{a-\mu}{\sigma}) - \phi(\frac{b-\mu}{\sigma})}{\Phi(\frac{a-\mu}{\sigma}) - \Phi(\frac{b-\mu}{\sigma})}\biggr)^2\biggr].
\end{align}
We evaluate the Fisher metric as $[g_a(\xi)]_{ij} = -\phi_a \expt_\xi(\partial_i\partial_j\Phi - h_{ij})$, where $h_{ij}$ is given by \eqref{hij1}--\eqref{hij3}. Without loss of generality, we set $h_{ij}=0$ and the Fisher metrics $(g_a)_{ij}$ and their first and second derivatives w.r.t. $\xi$, evaluated using \eqref{fisheralt}, can be written for the noise model \eqref{normL} with both the heteroscdasticity and positivity constraint as
\begin{align}
(g_a)_{ij} &= {\phi_a}\sum_{d=1}^{D}\Bigl[S_{i,d}^T\Sigma^{-1}_dS_{j,d} - \bigl(L_{d,ij}(\lambda(\alpha_d) * \sqrt{K_d})\bigr) + \frac{1}{2}\tr\bigl((\partial_i\partial_j\Sigma_d^{-1})J_d\bigr)\Bigr],\label{gsmod}\\
\begin{split}
\partial_k(g_a)_{ij} &= {\phi_a} \sum_{d=1}^{D}\Bigl[(\partial_k S_{i,d})^T\Sigma^{-1}_dS_{j,d} + S_{i,d}^T\Sigma^{-1}_d(\partial_kS_{j,d})\\ \qquad&+ S_{d,i}^T(\partial_k\Sigma_d^{-1})S_{d,j} - (\partial_k(\lambda(\alpha_d) * \sqrt{K_d})^T)\bigl(\Sigma^{-1}_d(\partial_iS_{d,j}) + (\partial_i\Sigma^{-1}_d)S_{d,j}\\ \qquad& +(\partial_j\Sigma^{-1}_d)S_{d,i}\bigr)-(\lambda(\alpha_d) * \sqrt{K_d})^T\bigl[(\partial_k\Sigma^{-1}_d)(\partial_iS_{d,i}) + \Sigma^{-1}_d(\partial_k\partial_iS_{d,i})\\\qquad& +(\partial_k\partial_i\Sigma^{-1}_d)^TS_{d,j} + (\partial_i\Sigma^{-1}_d(\partial_kS_{d,k}) + (\partial_k\partial_j\Sigma^{-1}_d)S_{d,i} + (\partial_j\Sigma^{-1}_d)(\partial_k S_{d,i})\bigr]\Bigr]\\ \qquad& + \frac{1}{2}\tr\Bigl((\partial_k\partial_j\partial_i\Sigma^{-1}_d)^TJ_d\bigr) + (\partial_i\partial_j\Sigma^{-1}_d)^T\partial_dJ_d\bigr)\Bigr),\label{dgsmod}
\end{split}
\end{align}
where
\begin{align}
L_{d,ij} &:= \Sigma_d^{-1}(\partial_iS_{d,j}) + (\partial_i\Sigma_d^{-1})S_{d,j} + (\partial_j\Sigma^{-1}_d)S_{d,i},\\
J_d &:= K_d *\bigl(\mathbf{1}_p - \alpha_d * \lambda(\alpha_d)\bigr).
\end{align}
Specifying the fixed and heteroscedastic noise components $\sigma$ and $\tilde{\sigma}$, the three terms $K_d$, $\lambda(\alpha_d)$ and $\alpha_d$ are defined by
\begin{equation}
\Sigma_d = \diag (K_d), \qquad
\alpha_d := X_d * K_d^{-1}, \qquad
\bigl[\lambda(\alpha_d)\bigr]_i := \frac{\phi(\alpha_{d,i})}{\Phi(\alpha_{d,i})},
\end{equation}
with $\Sigma_d$ defined in \eqref{sigmahet}, $X_d\equiv X_{\cdot, d}$, i.e. the $d^{\text{th}}$ column of $X$, and $\phi(\cdot)$ and $\Phi(\cdot)$ the density and cumulative distribution functions respectively of the standard normal\footnote{Not to be confused with the tempering parameter $\phi_a$ or the exponent $\Phi$ in the likelihood in \eqref{integrand}.}.

Removing the positivity constraints and setting the heteroscedastic variance term to zero, i.e. $a \rightarrow -\infty$ and $\tilde{\sigma}_d = 0$, we recover the expressions of the Fisher metric and its derivatives in the main text \eqref{gs} and \eqref{dgs}, respectively.

\subsection{Higher-order Ozaki discretisation of the Langevin diffusion}\label{Oz}
The Ozaki discretisation was formulated by Ozaki and Shoji \cite{Ozaki:1992tw, Shoji:1998da} and proposed for use in MALA framework by Stramer and Tweedie \cite{Stramer:1999wz}. In contrast to the Euler discretisation, it is more stable \cite{Roberts:2002hj}, providing a higher-order approximation for the drift term in the Langevin diffusion SDE. On the downside, its implementation can often be computationally expensive.

The Ozaki discretisation of the Langevin diffusion SDE can be expressed as an MVN proposal, like the Euler discretisation in \eqref{proposal}, as
\begin{equation}
q_a(\xi_{a+1}\mid\xi_a)\sim\mathcal{N}\bigl(\mu^O(\xi_a, \epsilon), \Sigma^O(\xi_a, \epsilon)\bigr),
\end{equation}
where the components of the sequence (as indexed by $a$) of means $\mu_a^O(\xi_a, \epsilon)$ and covariance matrices  $\Sigma_a^O(\xi, \epsilon)$ are given by
\begin{align}
\bigl(\mu_a^O(\xi_a, \epsilon)\bigr)^i &= (\xi_a)^i + \Bigl(J^{-1}_a\bigl(\exp(\epsilon^2 J_a) - \mathbb{I}_D\bigr)\Bigr)^i_{\hphantom{i}j}b^j,\label{omu}\\
\bigl(\Sigma_a^O(\xi_a, \epsilon)\bigr)^{ij} &= \frac{1}{2}g^{il}\Bigl(J^{-1}_a\bigl(\exp(2\epsilon^2 J_a) - \mathbb{I}_D\bigr)\Bigr)_{l}^{\hphantom{l}j},\label{osig}
\end{align}
where the drift term $b_a(\xi_a)$ is written as
\begin{equation}
b_a(\xi_a) = \frac{1}{2}g^{ij}\partial_j\ell - g^{ik}(\partial_jg_{kl})g^{lj}+ \frac{1}{2}g^{ij}g^{kl}\partial_jg_{kl},
\end{equation}
with the Jacobian $\bigl(J_a(\xi_a)\bigr)^i_{\hphantom{i}j} = \frac{\partial b_a^i}{\partial \xi_a^j}$. The explicit form of $J_a(\xi_a)$ is given by
\begin{equation}
\begin{split}
J_a(\xi_a)&=-\frac{1}{2}\bigl(g^{ik}(\partial_jg_{kl})g^{lm}\partial_m\ell + \frac{1}{2} \partial_j\partial_k\ell + g^{ik}(\partial_jg_{kl})g^{lm}(\partial_ng_{mp})g^{pn}\\
& \quad - g^{ik}(\partial_j\partial_l g_{km})g^{ml} + g^{ik} (\partial_ng_{kl})g^{lm}(\partial_j g_{mp}) g^{pn} -\frac{1}{2} g^{ik}(\partial_j g_{kl}) g^{lm}g^{np}(\partial_m g_pn)\\
& \quad + g^{ik}\bigl[ g^{lm} (\partial_j g_{mn}) g^{np} (\partial_k g_{pl}) + g^{lm} (\partial_j\partial_k g_{lm})\bigr]
\end{split}
\end{equation}

Expanding the first two terms of the exponentials in \eqref{omu} and \eqref{osig}, we have
\begin{align}
\bigl(\mu_a^O(\xi_a, \epsilon)\bigr)^i &\approx \xi_a^i + \epsilon^2 b_a^i + \frac{\epsilon^4}{2}(J_ab_a)^i, \\
\bigl(\Sigma_a^O(\xi_a, \epsilon)\bigr)^{ij} &\approx  g^{il}\bigl(\epsilon^2\mathbb{I}_D + 2\epsilon^4J_a\bigr)_l^{\hphantom{l}j}.
\end{align}
Keeping only the $O(\epsilon^2)$ terms of the expansion, we recover the Euler discretisation proposal parameters \eqref{eulerprop}.

\section*{Acknowledgements}
\noindent
AS and MPHS gratefully acknowledge financial support from EPSRC (EP/I017267/1); SF is funded through an MRC Computational Biology Research Fellowship; MPHS is a Royal Society Research Merit award holder.

\bibliographystyle{model1-num-names}
\bibliography{IGSMC_2}

\end{document}